\documentclass[epj]{svjour}
\usepackage{epsfig} 
\usepackage{psfig} 
\usepackage{times}

\begin{document}

\title{Near-threshold production of $a_0(980)$ mesons in the
reaction $pp \rightarrow d K^+ \bar{K}^0$}

\author{
V.~Yu.~Grishina\inst{1}, 
L.A.~Kondratyuk\inst{2},
M.~B\"uscher\inst{3}, 
W. Cassing\inst{4}} 

\institute{
Institute for Nuclear Research, 60th October Anniversary Prospect 7A, 117312
Moscow, Russia
\and 
Institute of Theoretical and Experimental Physics, B.\ Cheremushkinskaya 25,
117259 Moscow, Russia 
\and 
Institut f\"ur Kernphysik, Forschungszentrum J\"ulich,
D-52425 J\"ulich, Germany
\and
Institut f\"ur Theoretische Physik, Universit\"at Giessen, 
Heinrich-Buff-Ring 16, D-35392 Giessen, Germany
}

\date{Received: date / Revised version: date}

\abstract{Using an effective Lagrangian approach as well as the
Quark-Gluon Strings Model we analyze near-threshold production of
$a_0(980)$-mesons in the reaction  $NN \rightarrow d K \bar{K} $
as well as the background of non-resonant $K \bar{K}$-pair
production.  We argue that the reaction $pp \rightarrow d K^+
\bar{K}^0$ at an energy release $Q \leq 100$ MeV is dominated by
the intermediate production of the $a_0(980)$-resonance. At larger
energies the non-resonant $K^+ \bar{K}^0$-pair production --- where the
kaons are produced in a relative $P$-wave --- becomes important. The
effects of final-state interactions are evaluated in a unitarized
scattering-length approach and found to be in the order of a 20\%
suppression close to threshold. Thus in present experiments at the
Cooler Synchrotron COSY-J\"ulich for $Q \leq 107$ MeV the $a_0^+$
signal can reliably be separated from the non-resonant $K^+ \bar{K^0}$
background. }


\authorrunning{V.~Yu.~Grishina et al.}
\titlerunning{Near-threshold production of $a_0(980)$ mesons}

\maketitle

\section{Introduction}
During the last two decades the physics of the lightest scalar mesons
$a_0(980)$ and $f_0(980)$ has gained vivid attention. The constituent
quark model considers these scalar mesons as conventional $q \bar q$
states (see, e.g., Refs.
\cite{Morgan,Ani,Montanet,Anisovich,Narison} and references
therein), however, the structure of these states seems to be more
subtle. Alternative descriptions are $K\bar{K}$ molecules
\cite{Weinstein,Jan,Oller}, unitarized $q\bar{q}$ states
\cite{Tornqvist,Beveren} or  four-quark cryptoexotic states
\cite{Jaffe,Achasov,Vijande}.
 A further problem with these light scalar mesons is a possibly strong mixing
between the uncharged $a_0(980)$ and the $f_0(980)$ due to a
common coupling to $K\bar K$ intermediate states
\cite{Achasov,Achasov2,Barnes,Speth}. This effect will influence
the structure of the uncharged component of the $a_0(980)$ and implies
that a comparative study of the $a_0^0$ and $a_0^+$ (or $a_0^-$) has
to be performed. Moreover, the $a_0(980)$-$f_0(980)$ mixing can
generate isospin violation in different reactions with $a_0/f_0$
production \cite{Kud00,Kerbikov,Close2000,Grishina2001}.

At COSY-J\"ulich an experimental program on the study of
near-threshold $a_0/f_0$ production in $pp,~pn,~pd$ and $dd$
interactions has been started with the ANKE spectrometer
\cite{Buescher,Buescher1,COSY1,COSY2,COSY3}. Recently, first
results on the reaction $pp \rightarrow d K^+ {\bar K}^0$ near
threshold have become available at an excess energy of $Q=46$ MeV
\cite{Kleber}. The present study is devoted to the theoretical
analysis of these data. Furthermore, we provide predictions for
different observables at larger excess energy $Q$ and investigate the
influence of final-state interactions (FSI), the importance of which
has been pointed out in Ref.~\cite{Oset}.

In a recent work \cite{Grishina} we have considered $a_0$
production in the reactions $\pi N \rightarrow a_0 N$ and $NN
\rightarrow d a_0$ near threshold and at beam energies up to a few
GeV. An effective Lagrangian approach as well as the Regge-pole
model were applied to investigate different contributions to the
cross section of the reaction  $\pi N \rightarrow a_0N$. These
results were also used for an analysis of $a_0$ production in $NN$
collisions \cite{Brat02,Kondrat02}. In this paper we present a
more detailed study of the reaction $NN \rightarrow d K \bar{K} $
taking into account both the $a_0$ contribution to this reaction
as well as the non-resonant $K \bar K$ background. We demonstrate
that the $u$-channel mechanism --- normalised to the data from LBL
(Berkeley) for the reaction $pp\rightarrow d X$ at 3.8~GeV/c 
\cite{Abolins} --- can reproduce the total cross section of the reaction $pp
\rightarrow d a_0^+\rightarrow d K^+ \bar{K}^0$ at 3.46~GeV/c
($Q=46$~MeV) as measured at ANKE. However, it fails to reproduce the
distribution in the deuteron scattering angle. We show that
quantitatively better results can be achieved within the framework of
the Quark-Gluon Strings Model (QGSM).

Our paper is organized as follows: In Sect.~2 and 3 the two-step
model within the framework of an effective Lagrangian approach is
used for the analysis of different contributions for resonant
(through the $a_0$) and non-resonant production of $K \bar K$
pairs in the reaction $NN \rightarrow d K \bar{K} $. In Sect.~4
the reaction $NN \rightarrow da_0 \rightarrow d K \bar{K}$ is
considered additionally within the Quark-Gluon Strings Model while
in Sect.~5 a detailed analysis of final-state interactions (FSI)
is given. Our conclusions are presented in Sect.~6. The amplitudes
for the different contributions to the reactions $\pi N
\rightarrow a_0 N$ are given in the Appendix.

\section{Effective Lagrangian approach to the reaction
$N N \rightarrow d K \bar{K}$} \label{sec:da0}

\begin{figure}[t]
  \centerline{\psfig{file=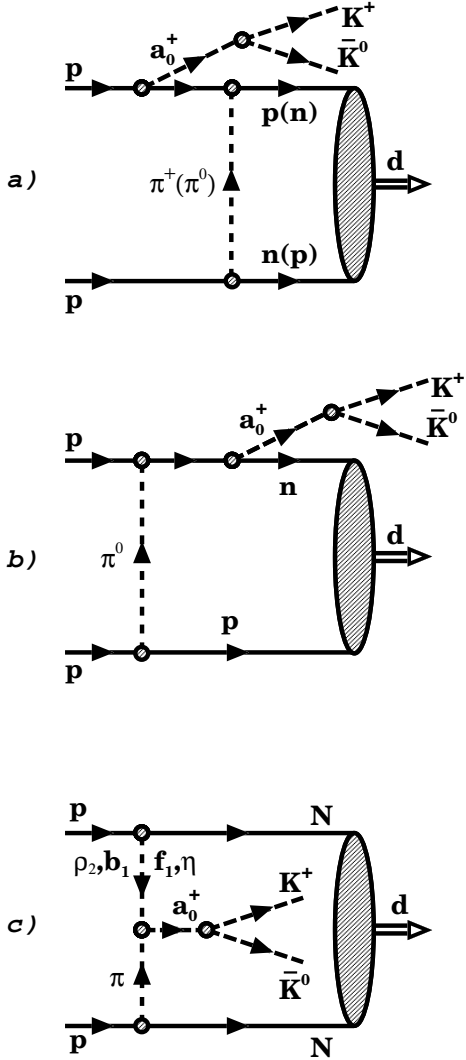,width=6.2cm}}
    \caption{Diagrams describing resonant
      contributions
    to the reaction $pp \rightarrow d K^+ \bar{K}^0$ within
   the framework of the two-step model.}
  \label{fig:tsm}
\end{figure}

\begin{figure}[t]
  \centerline{\psfig{file=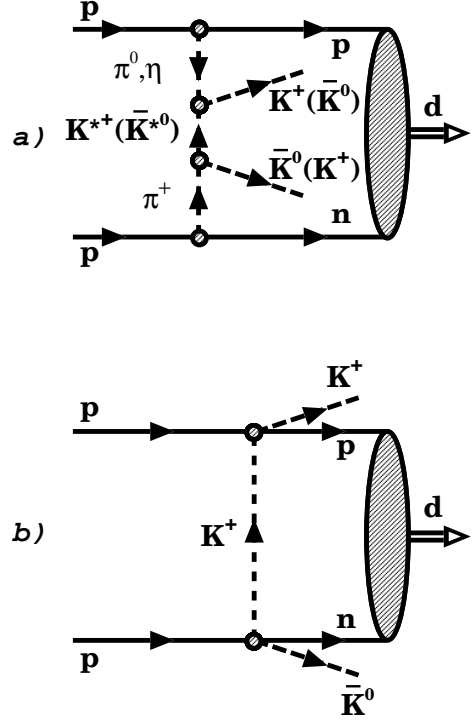,width=6.2cm}}
    \caption{Diagrams describing  non-resonant  mechanisms
    in the reaction $pp \rightarrow d K^+ \bar{K}^0$ within
    the framework of the two-step model.}
  \label{fig:tsm2}
\end{figure}

Within the framework of the two-step model (TSM) with one-pion
exchange in the intermediate state
(cf.~Refs.~\cite{Grishina1,Grishina2}) the contributions of
hadronic intermediate states to the amplitude of the reaction
$pp\rightarrow da_0^{+}\rightarrow d K^+ \bar{K^0}$ are described
by diagrams $a) - c)$ in Fig.~\ref{fig:tsm}. Accordingly, we
consider different contributions to the resonant amplitude $\pi N
\rightarrow a_0 N \rightarrow K \bar{K}N$:\\ i) the $u$- and
$s$-channel nucleon exchanges (Fig.~\ref{fig:tsm} a) and b),
respectively);
\\ ii) the $\eta$- and
$f_1(1285)$-meson exchanges (Fig.~\ref{fig:tsm} c);\\ iii) the
$b_1$ and $\rho_2$ Reggeon exchanges (Fig.~\ref{fig:tsm} c). \\
The non-resonant background contribution to the reaction $N N
\rightarrow d K \bar{K}$ is described by the diagrams  in
Fig.~\ref{fig:tsm2} a) and b) for $\pi-K^{\star}-\pi (\eta)$- and
$K$-exchange, respectively (see also Ref.~\cite{Kondrat02}).

Since we are interested in the $pp\rightarrow da_0^{+}$ and
$pp\rightarrow dK^+ \bar{K^0}$  cross sections near threshold,
where the momentum of the final deuteron is comparatively small,
we use a non-relativistic description of this particle by
neglecting the 4th component of its polarization vector.
Correspondingly, the relative motion of the nucleons in the
deuteron is also treated non-relativistically. The $pp\rightarrow
da_0^+$ and $pp\rightarrow dK^+\bar{K}^0$ amplitudes have to be
antisymmetrized with respect to permutation of the initial protons
$a$ and $b$ and therefore can be written as:
\begin{eqnarray}
T_{pp \rightarrow d a_0^+} (\mathbf{p}_a, \mathbf{q}_d)
&=& T_{pp\rightarrow d a_0^+}^{ab}(\mathbf{p}_a, \mathbf{q}_d)
\nonumber \\ &&
 - T_{pp\rightarrow da_0^+}^{ba}(\mathbf{p}_b,
\mathbf{q}_d) \ , \label{Tabbaa0} \\ 
T_{pp\rightarrow
dK^+\bar{K}^0} (\mathbf{p}_a, \mathbf{q}_d, \mathbf{q}_{12})
&=&T_{pp\rightarrow dK^+\bar{K}^0}^{ab}(\mathbf{p}_a, \mathbf{q}_d,
\mathbf{q}_{12}) \nonumber \\ &&
 -T_{pp\rightarrow dK^+ \bar{K}^0}^{ba}(\mathbf{p}_b,
\mathbf{q}_d, \mathbf{q}_{12}) \ . \label{Tabbakk}
\end{eqnarray}
Here and below the notations $q_1$, $q_2$, $q_d$, $p_a$ and $p_b$
are used for the 4-momenta of the $\bar{K}^0$, $K^+$, deuteron,
initial protons $a$ and $b$, respectively. We have introduced the
relative 3-momentum
$\mathbf{q}_{12}=(\mathbf{q}_1-\mathbf{q}_2)/2$ for the final
kaons, which are also considered as nonrelativistic particles for
excess energies
$Q\le 100 \div 150$~MeV. The motion of the nucleons $a^{\prime}$
and $b^{\prime}$ in the deuteron is described by the relative
momentum $\mathbf{p}_{b^{\prime}a^{\prime}} \equiv
(\mathbf{p}_{b^{\prime}}-\mathbf{p}_{a^{\prime}})/2=
\mathbf{p}_{b^{\prime}}-\mathbf{q}_d/2$. Then one can write the
first terms \newline
$T_{pp\rightarrow da_0^+}^{ab} (\mathbf{p}_a,
\mathbf{q}_d)$ and $T_{pp\rightarrow dK^+\bar{K}^0}^{ab}
(\mathbf{p}_a, \mathbf{q}_d,\mathbf{q}_{12})$ on the r.h.s. of
Eqs.~(\ref{Tabbaa0}) and~(\ref{Tabbakk}) as (\cite{Grishina1})
\begin{eqnarray}
&&\hspace*{-4mm} T_{pp\rightarrow d a_0^+}^{ab} (\mathbf{p}_a,
\mathbf{q}_d)
 = \frac{f_{\pi NN}}{m_\pi} \ (p^0+m_N) \ (2 m_N)^{3/2}
\label{T_aba0} \nonumber \\ && \times \displaystyle\sum _{X(a_0)}
M^{\{X(a_0)\}\,jl}_{pp\rightarrow d a_0^+}({\mathbf p}_a, {\mathbf
q}_d)\ \varphi_{\lambda_a}^T (\mathbf {p}_a)\nonumber \\ &&
 \times (-i\sigma_2)\sigma^j \mbox{\boldmath $\sigma$} \cdot
\mbox{\boldmath{$\epsilon$}}^{*(d)} \sigma^l
\varphi_{\lambda_b}(\mathbf{p}_b)\ ,
 \\ &&\hspace*{-4mm}
T_{pp\rightarrow dK^+\bar{K}^0}^{ab} (\mathbf{p}_a, \mathbf{q}_d,
\mathbf{q}_{12})
 = \frac{f_{\pi NN}}{m_\pi} \ (p^0+m_N) \ (2 m_N)^{3/2}
\label{T_abkk} \nonumber \\ && \times \displaystyle\sum _{X}
M^{\{X\}\,jl}_{pp\rightarrow dK^+\bar{K}^0} ({\mathbf p}_a,
{\mathbf q}_d, {\mathbf q}_{12})\ \varphi_{\lambda_a}^T (\mathbf
{p}_a) \nonumber \\ &&  \times (-i\sigma_2)\sigma^j
\mbox{\boldmath $\sigma$} \cdot
\mbox{\boldmath{$\epsilon$}}^{*(d)} \sigma^l
\varphi_{\lambda_b}(\mathbf{p}_b)\ ,
\end{eqnarray}
where $\mathbf{p}_a=-\mathbf{p}_b=\mathbf{p}$, $p^0=
p_a^0=p_b^0=\sqrt{\mathbf{p}^2+m_N^2}$ in the center-of-mass
frame. The tensor functions  ${M}^{\{X(a_0)\}\, jl}_{pp\rightarrow
d a_0^+}(\mathbf{p}_a, \mathbf{q}_d)$ and ${M}^{\{X\}\,
jl}_{pp\rightarrow dK^+\bar{K}^0}(\mathbf{p}_a, \mathbf{q}_d,
\mathbf{q}_{12})$ are defined by the integrals
\begin{eqnarray}
&&\hspace*{-4mm} M^{\{X(a_0)\}\, jl}_{pp\rightarrow d a_0^+}
(\mathbf{p}_a, \mathbf{q}_d) = \int
\frac{d^3p_{b^{\prime}a^{\prime}}}{(2\pi)^{\, 3/2}} \
\Psi_d(\mathbf{p}_{b^{\prime}a^{\prime}}) \label{Mnrinta0} \nonumber \\
&\times& \left\{-\frac{p_a^j}{p^0+m_N} + \frac{(-2\,
p_{b^{\prime}a^{\prime}}+q_d)^{j}}{4\, m_N}\right\} \nonumber \\
&\times&  \Phi_{\pi N\rightarrow a_0 N}^{\{X(a_0)\}\, l}
(\mathbf{p}_a, \mathbf{q}_d, \mathbf{p}_{b^{\prime}a^{\prime}})\
\frac{F_{\pi NN}(t_{a a^{\prime}})}{t_{a a^{\prime}}-m_\pi ^2}\ ,
 \\ &&\hspace*{-4mm} M^{\{X\}\, jl}_{pp\rightarrow
dK^+\bar{K}^0} (\mathbf{p}_a, \mathbf{q}_d, \mathbf{q}_{12}) =
\int \frac{d^3p_{b^{\prime}a^{\prime}}}{(2\pi)^{\, 3/2}} \
\Psi_d(\mathbf{p}_{b^{\prime}a^{\prime}}) \label{Mnrintkk} \nonumber \\
&\times& \left\{-\frac{p_a^j}{p^0+m_N} + \frac{(-2\,
p_{b^{\prime}a^{\prime}}+q_d)^{j}}{4\, m_N}\right\} \nonumber \\
&\times& \Phi_{\pi N \rightarrow K \bar{K} N}^{\{X\}\, l}
(\mathbf{p}_a, \mathbf{q}_d, \mathbf{q}_{12},
\mathbf{p}_{b^{\prime}a^{\prime}})\ \frac{F_{\pi NN}(t_{a
a^{\prime}})}{t_{a a^{\prime}}-m_\pi ^2}\ .
\end{eqnarray}
Here $\Psi_d(\mathbf{p}_{b^{\prime}a^{\prime}})$ is the deuteron wave function,
 $t_{aa^{\prime}}=(p_a-p_{a^{\prime}})^2$ is
the virtual pion momentum squared. The vector functions
$$\Phi_{\pi N\rightarrow a_0 N}^{\{X(a_0)\}\, l} (\mathbf{p}_a,
\mathbf{q}_d, \mathbf{p}_{b^{\prime}a^{\prime}} )$$ and
$$\Phi_{\pi N\rightarrow K\bar{K} N}^{\{X \}\, l} (\mathbf{p}_a,
\mathbf{q}_d, \mathbf{q}_{12}, \mathbf{p}_{b^{\prime}a^{\prime}}
)$$ depend on the mechanisms~$X(a_0)$ (or $X$) of the $a_0$ (or
$K\bar{K}$) production. For each vertex with a virtual meson we
use the monopole form factor
\begin{eqnarray}
&&F_{j}(t)=\frac{\Lambda_j^2-m_j^2}{\Lambda_j^2-t} \ ,
\label{monFF}
\end{eqnarray}
where the $\Lambda_j$ denote a  cut-off parameter,
$\Lambda_{\pi}=1.3$~GeV.

In the case of the $K\bar{K}$ production
via $a_0$ resonance we have the well-known convolution formula
\begin{eqnarray}
&&\hspace*{-4mm} \Phi_{\pi N\rightarrow K \bar{K} N}^ {\{X(a_0)\}
\ l} (\mathbf{p}_a, \mathbf{q}_d, \mathbf{q}_{12},
\mathbf{p}_{b^{\prime}a^{\prime}}) =  \Phi_{\pi
N\rightarrow a_0 N}^ {\{X(a_0)\} \ l} (\mathbf{p}_a,
\mathbf{q}_d,\mathbf{p}_{b^{\prime}a^{\prime}}) \nonumber \\ &&
\times
 F_0(m_{a_0})
\label{flatte}
\end{eqnarray}
where $F_0(m_{a_0})$ is the Flatt\'e mass distribution amplitude
(see, e.g. Ref.\cite{Abele}), $m_{a_0}=\sqrt{(q_1+q_2)^2}$ and
\begin{eqnarray}
&&\hspace*{-4mm} \Phi_{\pi N\rightarrow K^+ \bar{K}^0 N}^
{\{X(a_0)\} \ l} (\mathbf{p}_a, \mathbf{q}_d,
\mathbf{p}_{b^{\prime}a^{\prime}}) = \label{respi} \nonumber \\ &&
I^{\{X(a_0)\}}\ \left\{ \left[ \vphantom
{p_{b}^{l}\left(\frac{\displaystyle q_{a_0}^0+m_N+
\frac{\mathbf{p}_{b^{\prime}a^{\prime}}\cdot
\mathbf{q}_{d}}{2m_N}}{p^0+m_N}\right)}
 -\frac{p_b^l}{p^0+m_N} +
\frac{(2\, p_{b^{\prime}a^{\prime}}+q_d)^{l}}{4\, m_N}\right]
\right. \nonumber \\ &&
 \times A^{\{X(a_0)\}}(s_{\{a_0, \, b^{\prime}\}}, t_{bb^{\prime}})
 \nonumber
\\ &+& \left[p_{b}^{l}\left(\frac{\displaystyle q_{a_0}^0+m_N+
\frac{\mathbf{p}_{b^{\prime}a^{\prime}}\cdot
\mathbf{q}_{d}}{2m_N}}{p^0+m_N}\right) \right. \nonumber \\ &&  +
p_{b^{\prime}a^{\prime}}^{l}\left(\frac{\displaystyle
q_{a_0}^0-m_N+ \frac{\mathbf{p}_{b}\cdot
\mathbf{q}_{d}}{p^0+m_N}}{2m_N}\right)
 \nonumber \\
&&\left.  + q_{d}^{l}\left(\frac{\displaystyle
q_{a_0}^0+3m_N- \frac{\mathbf{p}_{b} \cdot
\mathbf{p}_{b^\prime a^\prime}}{p^0+m_N}}{4m_N}\right) \right]
\nonumber \\
&& \left. \times
 B^{\{X(a_0)\}}(s_{\{a_0, \, b^{\prime}\}}, t_{bb^{\prime}}) \
\right\} .
\end{eqnarray}
Here $I^{\{X(a_0)\}}$ denotes the isospin factor, \begin{equation}
s_{\{a_0,\,
b^{\prime}\}}=(q_{a_0}+p_{b^{\prime}})^2, \hspace{0.5cm}
t_{bb^{\prime}}=(p_b-p_{b^{\prime}})^2 \nonumber \end{equation}
 and the 4-momentum of
$a_0$ is defined as $q_{a_0}=p_a+p_b-q_d$. Two invariant
amplitudes \begin{equation} A^{\{X(a_0)\}}(s_{\{a_0, \,
b^{\prime}\}}, t_{bb^{\prime}}) \end{equation} and
\begin{equation} B^{\{X(a_0)\}}(s_{\{a_0, \, b^{\prime}\}}, t_{bb^{\prime}})
\end{equation}
define the $s$-channel helicity amplitudes for the $\pi N
\rightarrow a_0 N$ reaction as follows~\cite{Achasov2}
\begin{eqnarray} \label{helamp}
&& M_{\lambda_{b^{\prime}}\lambda_{b}}({\pi^- p \rightarrow a_0
N})=  \nonumber \\ &&
\bar{u}_{\lambda_{b^{\prime}}} \gamma_{5} \left\{-A(s, t)
  -\frac{1}{2}\,\gamma ^{\mu}\left
(q_{\pi}+q_{a_0}\right )_{\mu} B(s, t) \right\}u_{\lambda_{b}} .
\end{eqnarray}
The amplitudes for different mechanisms of the
$\pi^{-}p\rightarrow a_{0}N$ reactions are given in the Appendix
for completeness. In the case of the $s$-, $u$-channel nucleon
exchanges as well as $\rho_2$-, $b_1$-Reggeon exchanges we fix the
parameters of the invariant amplitudes $A(s,t)$ and $B(s,t)$ using
the $\pi^{-} p\rightarrow a_0^0 n$ channel. Since the isoscalar
$\eta$ and $f_1$ exchange mechanisms do not contribute to this
reaction we choose the $\pi^{-} p\rightarrow a_0^- p$ channel to
define parameters of the amplitudes $A(s,t)$ and $B(s,t)$. Then we
can fix the isospin coefficients for different mechanisms in Eq.
(\ref{respi}) as follows: $I^{\{u\}}=3$, $I^{\{s\}}=1$,
$I^{\{{\rho}_2\}}=I^{\{b_1\}}=2$,
$I^{\{\eta\}}=I^{\{f_1(1285)\}}=\sqrt{2}$.

The non-resonant $K \bar{K}$ production via $K^{\star} - P $-
exchange with a pseudoscalar meson $P=\pi^{0}$ or $\eta$ is
given by
\begin{eqnarray}
&&\hspace*{-4mm} \Phi_{\pi N\rightarrow K \bar{K} N}^ {\{K^{\star}
- P\} \ l} (\mathbf{p}_a, \mathbf{q}_d, \mathbf{q}_{12},
\mathbf{p}_{b^{\prime}a^{\prime}}) =\label{Nrespi} \nonumber \\ && \frac{F_{P
NN}(t_{b b^{\prime}})}{t_{b b^{\prime}}-m_{P}^2} \sqrt{2} \
T_{\pi^{+} P\rightarrow K^+\bar{K}^0}(\mathbf{p}_a, \mathbf{q}_d,
\mathbf{q}_{12}, \mathbf{p}_{b^{\prime}a^{\prime}}) \nonumber \\
&\times & \left\{-\frac{p_b^l}{p^0+m_N} + \frac{(2\,
p_{b^{\prime}a^{\prime}}+q_d)^{l}}{4\, m_N}\right\} ,
\end{eqnarray}
where the elementary $\pi^{+} P \rightarrow K^{+} \bar{K}_0$
transition amplitude has the form
\begin{eqnarray}
&&\hspace*{-6mm} T_{\pi^{+} P\rightarrow
K^+\bar{K}^0}(\mathbf{p}_a, \mathbf{q}_d, \mathbf{q}_{12},
\mathbf{p}_{b^{\prime}a^{\prime}})= g_{K^{\star} \pi K}\
g_{K^{\star} P K} \sqrt{2} \label{Kstarplus} \nonumber
\\
&\times&
\left\{(p_a-p_{a^{\prime}}+q_1)_{\mu}(p_b-p_{b^{\prime}}+q_2)^{\mu}
\frac{(t_{aa^{\prime}}-m_{K}^2)(t_{bb^{\prime}}-m_{K}^2)}{m_{K^{\star\,
2}}} \right\} \nonumber\\ &\times& \frac{F_{ \pi K
K^{\star}}(t_{aa^{\prime}})\ F_{K^{\star} \pi K}(t_{K^{\star}})\
F_{K^{\star} P K}(t_{K^{\star}})\  F_{P K
K^{\star}}(t_{bb^{\prime}})} {t_{K^{\star}}-m_{K^{\star \, 2}}}\ .
\end{eqnarray}
Here $t_{K^{\star}}=(p_a-p_b-p_{a^{\prime}}+p_{b^{\prime}})^2$.
The coupling constants $ g_{K^{\star} \pi K}=-3.02$, $g_{K^{\star}
\eta K}= \sqrt{3}\ g_{K^{\star} \pi K}$ and the cut-off parameter
for the virtual $K^{\star}$ exchange
$\Lambda_{K^{\star}}(K^{\star} \eta K)=3.29$ GeV are taken from
Ref. \cite{Jan}. The remaining cut-off parameter
$\Lambda_{K^{\star}}(K^{\star} \pi K)$ is adjusted to reproduce
the experimental data \cite{Kleber} (see Sect. 3). We note that
the amplitude~(\ref{Kstarplus}) takes into account only the
$K^{\star +}$-exchange. In the case of the~$P=\pi (\eta)$ we
should subtract (add) the corresponding
$\bar{K}^{\star0}$-exchange amplitude (obtained by the
substitution~$q_1\leftrightarrow q_2$ in Eq.~(\ref{Kstarplus})).
This rule follows from $G$-parity conservation. We recall that the
$G$-parity of the $K\bar{K}$-system with orbital momentum~$L$ and
isospin~$I$ is given by~$(-1)^{L+I}$. Therefore, for~$I=1$ in our
case the orbital momentum of the~$K\bar{K}$-pair should be odd for
positive $G$-parity and even for negative $G$-parity. Thus the
non-resonant $S$-, $D$-\ldots wave $K\bar{K}$-pair production in the
$pp \rightarrow d K^{+} \bar{K}^0$ reaction is contributed by the
$\pi - K^{\star} -\eta$-exchange mechanism (see also Sect.~3). The
non-resonant $\pi - K^{\star} -\pi$-exchange part of the $pp
\rightarrow d K^{+} \bar{K}^0$ amplitude near threshold leads
to~$P$, $F$-\ldots wave $K\bar{K}$-pair production.

For the sake of completeness we have calculated also the
$K$-exchange term defined by the diagram of ~(Fig. \ref{fig:tsm}
e)). The corresponding amplitude reads
\begin{eqnarray}
&&\hspace*{-1mm} T^{\{K\}ab}_{pp\rightarrow dK^+\bar{K}^0}
(\mathbf{p}_a, \mathbf{q}_d, \mathbf{q}_{12})
 = \frac{1}{\sqrt{2 m_N}}  \times
\label{T_ab_k} \\ && M^{\{K\}}_{pp\rightarrow
dK^+\bar{K}^0}({\mathbf p}_a, {\mathbf q}_d, {\mathbf q}_{12})\
\varphi_{\lambda_a}^T (\mathbf {p}_a)\ (-i\sigma_2)
\mbox{\boldmath $\sigma$} \cdot
\mbox{\boldmath{$\epsilon$}}^{*(d)}
\varphi_{\lambda_b}(\mathbf{p}_b) \nonumber
\end{eqnarray}
with the scalar function
\begin{eqnarray}
&&\hspace*{-4mm} M^{\{K\}}_{pp\rightarrow dK^+\bar{K}^0}
(\mathbf{p}_a, \mathbf{q}_d, \mathbf{q}_{12}) = \int
\frac{d^3p_{b^{\prime}a^{\prime}}}{(2\pi)^{\, 3/2}} \
\Psi_d(\mathbf{p}_{b^{\prime}a^{\prime}}) \label{Mnrintk} \nonumber \\
&\times&A_{KN\rightarrow KN}(\mathbf{p}_a, \mathbf{q}_d,
\mathbf{q}_{12}) \ A_{\bar{K}N\rightarrow \bar{K}N}(\mathbf{p}_a,
\mathbf{q}_d, \mathbf{q}_{12}) \nonumber \\ &\times&
\frac{F_{KNN}^{2}(t_{K})}{t_{K}-m_K^2} \ .
\end{eqnarray}
Here $t_K$ is the squared 4-momentum of the virtual kaon. For
the~$KN$(~$\bar{K} N$) cross sections we used the parametrizations
from Ref.\cite{Cugnon}. The cut-off parameter $\Lambda_K$ was
taken to be 1.2 GeV (see, e.g. Ref.\cite{Sibirtsev1}).

Keeping in mind that the nucleons in the deuteron are considered
as nonrelativistic particles, the momentum transfers
\newline squared in the denominators of the propagators in
Eqs.~(\ref{Mnrinta0},\ref{respi}) can be rewritten as follows
\begin{eqnarray}
t_{a a^{\prime}}&\simeq& -2\left( p^0-m_N \right) \ m_N-\frac{p^0}{m_N}\
\left(-\mathbf{p}_{b^{\prime}a^{\prime}}+\frac{\mathbf{q}_d}{2}\right)^2
\nonumber \\ &&
-2\mathbf{p}_a \cdot \mathbf{p}_{\, b^{\prime}a^{\prime}}
+\mathbf{p}_a \cdot \mathbf{q}_d \ , \nonumber \\
t_{b b^{\prime}}&\simeq& -2\left( p^0-m_N \right) \ m_N-\frac{p^0}{m_N}\
\left(\mathbf{p}_{b^{\prime}a^{\prime}}+\frac{\mathbf{q}_d}{2}\right)^2
\nonumber \\ &&
-2\mathbf{p}_a \cdot \mathbf{p}_{\, b^{\prime}a^{\prime}}
-\mathbf{p}_a \cdot \mathbf{q}_d \ , \nonumber \\
t_{K^{\star}}&\simeq& t_K \simeq -\left(\mathbf{p}_a + \mathbf{p}_{\, b^{\prime}a^{\prime}}
-\mathbf{q}_{12}\right)^2 \ .
 \end{eqnarray}

The structure of the amplitudes (\ref{Tabbaa0}) and
(\ref{Tabbakk}) guarantees that their $S$-wave parts (when the initial
and final states have orbital momenta equal to zero) vanish since
they are forbidden by angular momentum conservation and the Pauli
principle. The second terms $T_{pp\rightarrow d
a_0^{+}}^{ba} (\mathbf{p}_b, \mathbf{q}_d)$ and $T_{pp\rightarrow
dK^+\bar{K}^0}^{ba}(\mathbf{p}_b, \mathbf{q}_d, \mathbf{q}_{12})$ on
the r.h.s.\ of Eqs.~(\ref{Tabbaa0}) and (\ref{Tabbakk}) can be
obtained from the first ones~$T_{pp\rightarrow d a_0^+}^{ab}
(\mathbf{p}_a, \mathbf{q}_d)$ (\ref{T_aba0}) and $T_{pp\rightarrow
dK^+\overline{K}^0}^{ab} (\mathbf{p}_a, \mathbf{q}_d,\mathbf{q}_{12})$
(\ref{T_abkk}) by exchanging $p_a \leftrightarrow p_b$.

\section{$a_0$ cross section and non-resonant background in the
reaction $pp \rightarrow d K^+ \bar{K^0}$}

\subsection{$a_0$-resonance contribution}

\begin{figure}[t]
  \centerline{\psfig{file=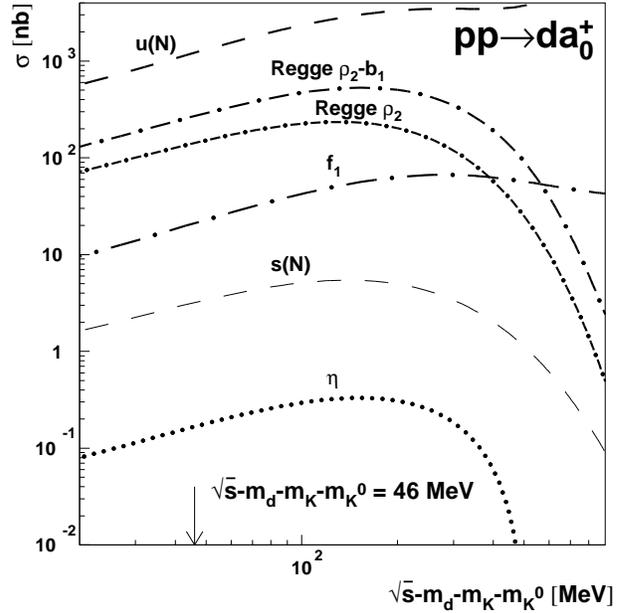,width=8cm}}
    \caption{Total cross section of the $pp \rightarrow d a_0^+$
    reaction as a function of the c.m. excess energy. 
    The contributions of the $u$- and $s$-channel
exchanges are shown by the bold dashed and thin dashed lines,
respectively. The lower long-dashed-dotted line and the dotted
line describe the $f_1$- and $\eta$- exchanges. The dash-dotted
line stands for the combined $\rho_2$ and $b_1$ Reggeon exchanges,
while the model result for the single $\rho_2$ Reggeon exchange is
shown by the short-dashed-dotted line. The arrow indicates the
the excess energy  $Q$=46 MeV of the ANKE experiment. }
  \label{fig:siga0tot1}
\end{figure}

To illustrate the hierarchy of the different mechanisms in the
case of $a_0$ production we present in Fig.~\ref{fig:siga0tot1}
our results for the total cross section of the reaction $pp
\rightarrow d a_0^+$. As in Ref.~\cite{Grishina} the $a_0NN$
coupling constant was taken from the Bonn model \cite{Holinde}.  For
the virtual nucleon we used the standard form factor given by
Eq.~({\ref{FN}}) in the Appendix with a cut-off parameter
$\Lambda_{N}=1.3$~GeV, which satisfies the constraints found in our
recent analysis of the $\pi N \rightarrow N K \bar{K}$ and $NN
\rightarrow NN K\bar{K}$ reactions~\cite{Kondrat02} (see comment after
Eq.~({\ref{FN}})).  Moreover, using this approach we can
simultaneously describe the LBL data on the forward differential cross
section of the reaction $pp \rightarrow d a_0^+$ at 3.8 GeV/c
\cite{Abolins}. In practical terms: the cut-off parameter $\Lambda_N$
may also be defined by normalizing the $u$-channel contribution to the
LBL data.

The parameters of the Regge model have been fixed by Achasov and
Shestakov \cite{Achasov2} in fitting Brookhaven data on the
reaction $\pi^- p \rightarrow a_0 ^0 n$ at 19 GeV/c
\cite{Dzierba}. All other parameters were taken the same as in
Ref.\cite{Kondrat02} (see also Appendix). As seen in
Fig.~\ref{fig:siga0tot1} the dominant contribution to the cross
section of the reaction $pp \rightarrow d a_0^+$ near threshold
comes from the $u$-channel mechanism (shown by the bold dashed
line) and all other contributions from $f_1$- and $\eta$-meson
exchanges, $s$-channel nucleon exchange and $b_1$- and $\rho_2$-
Reggeons can be neglected (for the forward differential cross
section this result was obtained earlier in Ref.\cite{Grishina}).

The $a_0$-resonance contribution to the cross section of the reaction
$pp \rightarrow d K^+ \bar{K^0}$ is calculated by convoluting the
cross section of the $a_0^+$ production with the Flatt\'e mass
distribution (see Eq.(8) and also Ref.\cite{Kondrat02}). The result
for the dominant $a_0$-resonance part corresponding to the diagram in
Fig.~\ref{fig:tsm}~a) is shown by the long-dashed line in Fig.~4. The
parameters of the Flatt\'e mass distribution are taken from
Ref.\cite{Abele}: $m_0=999$ MeV, $g_{\pi \eta}=324$ MeV and $g^2_{K
\bar K}/g^2_{\pi \eta}=$ 1.03. As it follows from Fig.~3 the total
cross section of the reaction $pp \rightarrow d a_0^+$ at
$p_{\mathrm{lab}}=3.46$ GeV ($Q=46$ MeV) in the narrow $a_0$ width
limit is about 1.2 $\mu$b. After convolution with the Flatt\'e
distribution we find that $\sigma(pp \rightarrow d a_0^+ \rightarrow
K^+ \bar{K}^0)$ is about 28~nb (see Fig.~4). The effective branching
ratio for the $a_0$ decay to the $K \bar K$ mode is 0.023 at $Q=46$
MeV. Such a large suppression as compared with the standard value
$\Gamma_{K\bar K}/\Gamma_{\pi \eta}= 0.177\pm0.024$~\cite{PDG} is
related to the phase space limitation and the $P$-wave character of
$a_0$ production in the reaction $pp \rightarrow d a_0^+$ near
threshold.

\subsection{Background contributions}

\begin{figure}[t]
  \centerline{\psfig{file=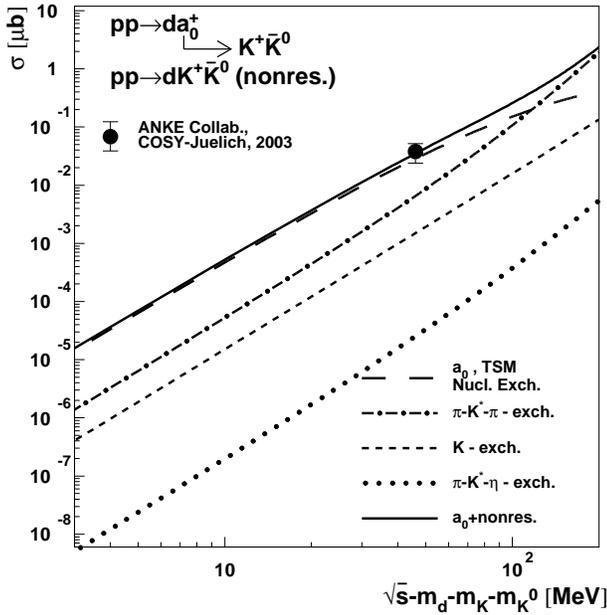,width=8cm}}
    \caption{Total cross section of the $pp \rightarrow d K^+ \bar{K}^0$
    reaction as a function of the c.m.\ excess energy. The $a_0$-resonance
    part of the cross section
    is displayed by the long-dashed line.
    The dash-dotted   and dotted lines show the background
corresponding to $\pi-K^{\star}-\pi$- and $\pi-K^{\star}-\eta$-
exchange mechanisms, respectively. The $K$-exchange contribution
is shown by the short-dashed line. The solid line displays the sum
of the all contributions. The bold point shows the experimental
cross section from Ref. \cite{Kleber}.}
  \label{fig:siga0totQGSM}
\end{figure}

An important problem is to understand the role of the non-resonant
contribution to the $pp \rightarrow dK^+ {\bar K}^0$ cross section. In
Ref. \cite{Kondrat02} the $\pi-K^{\star}-\pi(\eta)$-exchange
mechanisms for non-resonant $K\bar{K}$ production in the reactions
$\pi N\rightarrow N K\bar{K}$ and $NN \rightarrow NNK {\bar K}$ has
been considered. The results of calculations for the $\pi
N\rightarrow$ $ NK {\bar K}$ cross sections in different isospin
channels showed that the $a_0$-resonant part is expected to be more
pronounced at $Q\leq 250$~MeV while the non-resonant background might
become dominant at $Q\geq 250$~MeV (see Fig.~4 in
Ref.~\cite{Kondrat02}). The analysis of different isospin channels of
the reaction $NN \rightarrow NN K \bar K$ demonstrated that the
production of the $a_0$ --- as compared to the background --- is more
pronounced in the reaction $pp \rightarrow pn K^+ \bar{K}^0 $ than in
the reaction $pp \rightarrow pp K^+ {K}^-$.

Here we use these previous results to analyze the role of the
non-resonant background in the $pp \rightarrow d K^+ \bar{K}^0$
reaction. The diagrams describing $\pi-K^{\star}-\pi(\eta)$- and
$K$-exchange mechanisms are shown in Fig.~\ref{fig:tsm2} a) and b),
respectively. The results of the calculations are presented in
Fig.~4. The dash-dotted and dotted lines in
Fig.~4 display the background corresponding to
$\pi-K^{\star}-\pi$- and $\pi-K^{\star}-\eta$- exchange
mechanisms, respectively, while the $K$-exchange contribution is
shown by the short-dashed line. It can be seen from
Fig.~4 that this contribution is much smaller
than the cross section for the $\pi-K^{\star}-\pi$-exchange and
may savely be neglected.

As follows from the $G$-parity constraints (see comment after
Eq.~(15)) the $\pi-K^{\star}-\pi$ mechanism contributes mainly to
the $P$-wave in the $K^+ \bar{K}^0$-system, while the
$\pi-K^{\star}-\eta$-mechanism contributes dominantly to the
$S$-wave. The latter, in principle, via $K \bar K$-FSI can
contribute to the resonant $a_0$ channel where the kaons are
also produced in a relative $S$-wave. However, we neglect this 
in the following since the contriburion from this channel is 
very small (see dotted line in Fig.~4) and conclude that $K \bar K$
pairs from background will predominantly be in a $P$-wave, while in
the case of $a_0$ decay it will be produced in the $S$-wave (see
also Section~2 and Ref.\cite{Kondrat02}). According to the
long-dashed line in Fig.~4 the resonant part is
dominant up to $Q\simeq 100$ MeV. The background is seen to give
an important contribution only for $Q \geq 100$ MeV.

As mentioned before, the TSM gives an integrated cross section of
about 28~nb at $Q=46$~MeV for the $a_0$ resonance part. As
concerning the contribution of the $P$-wave  $ K \bar{K}$ pairs, 
we normalized it here to 6.5 nb at the same $Q$. This
value was obtained in Ref.~\cite{Kleber} from the best fit to the
data. To describe it within the $\pi -K^{\star}- \pi$-exchange
model we use a the cut-off parameter
$\Lambda_{K^{\star}}(K^{\star} \pi K)=1.25$ GeV. Using
Eqs.~(\ref{Tabbakk}), (\ref{T_abkk}), (\ref{Mnrintkk}) and
(\ref{Nrespi})--(\ref{Kstarplus}) one can find that the leading
term for the $ K \bar{K}$ $P$-wave part of the $pp \rightarrow d
K^+\bar{K^0}$ amplitude has the following spin structure
\begin{eqnarray}
T_{pp\rightarrow dK^{+} \bar{K^0}}^{{\pi-K^{\star}-\pi}} &\sim&
\varphi_{\lambda_a}^T (\mathbf {p}_a)\ (-i\sigma_2)
(\mbox{\boldmath $\sigma$} \cdot \mathbf {p}_a) (\mbox{\boldmath
$\sigma$} \cdot \mbox{\boldmath{$\epsilon$}}^{*(d)})\times
\nonumber \\ &&(
\mbox{\boldmath $\sigma$} \cdot \mathbf {p}_a)
 (\mathbf {p}_a \cdot \mathbf {q}_{12}) \
\varphi_{\lambda_b}(\mathbf{p}_b) \ .
\end{eqnarray}
Therefore, within the $\pi-K^{\star}-\pi$-exchange
model the background has the following angular distribution
\begin{eqnarray}
&&\frac{d\sigma}{d \Omega_{12}} \simeq N \cos^{2} \theta_{12} \ ,
\label{Omega12}
\end{eqnarray}
where $d \Omega_{12}$ = $d \cos \theta_{12}\  d \varphi_{12}$ with
$\Omega_{12}$ being the solid angle for the $K \bar K$ relative
momentum~$\mathbf{q}_{12}$. The angular distribution in
$\theta_{12}$ as given by Eq.~(\ref{Omega12}) is in a good
agreement with the experimental data \cite{Kleber}. However, the
TSM does not describe the distribution on the deuteron scattering
angle: it predicts a forward peak \cite{Grishina} instead of a
forward dip found in the ANKE experiment (see Fig.~4 in
Ref.~\cite{Kleber}). A possible solution of this discrepancy is
presented in the next Section within the Quark-Gluon Strings Model (QGSM).

\section{The reaction $NN \rightarrow d a_0$ in the QGSM}

\begin{figure}[t]
\centerline{\psfig{file=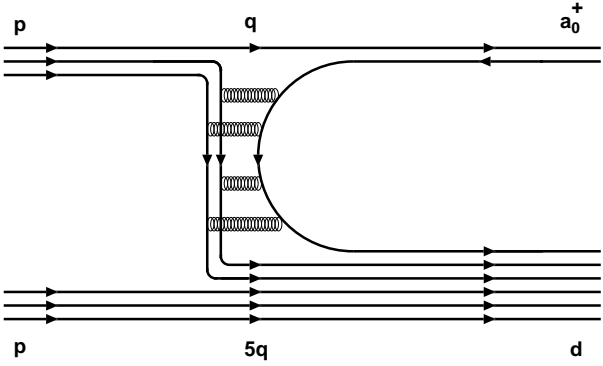,width=8cm}} \vspace{0.5cm}
\caption{Planar quark diagram describing the reaction $pp
\rightarrow a_0^+ d$ in the quark-gluon-strings model (QGSM).}
\label{fig:qgsm}
\end{figure}

As we have argued in the previous section the model based on the
effective Lagrangian approach can describe the energy behaviour of
the total cross section of the reaction $NN \rightarrow d a_0$.
However, it fails to reproduce the angular dependence of the
differential cross section. Remarkably, even at threshold the
typical values of the momentum transfer in the reaction $NN
\rightarrow d a_0$ exceed 1 GeV$^2$. Thus a complete description
of this reaction would require to take into account relativistic
effects as well as quark degrees of freedom. This can be done, for
example, within the framework of the Quark-Gluon Strings Model
(QGSM), which recently has successfully been applied in
Refs.~\cite{GrishinaGDPN1,KondratGDPN,GrishinaGDPN2} to the
description of deuteron photodisintegration at energies above
1~GeV at all angles.

This model --- proposed originally  by Kaidalov
\cite{Kaidalov,Kaidalov99} --- is based on two ingredients: i) a
topological expansion in QCD and ii) a space-time picture of the
interactions between hadrons that takes into account the
confinement of quarks. In a more general sense the QGSM can be
considered as a microscopic (nonperturbative) model of Regge
phenomenology for the analy\-sis of exclusive and inclusive
hadron-hadron and photon-hadron reactions on the quark level. The
main assumption of the QGSM is that the amplitudes $T(\gamma d
\rightarrow pn)$  and $T(NN \rightarrow a_0 d)$ can be described
by planar graphs with three valence-quark exchange in $t$ (or
$u$)-channels with any number of gluon exchanges between them
(Fig.\ref{fig:qgsm}). This corresponds to the contributions of the
$t$- and $u$-channel nucleon Regge trajectories. In the space-time
picture the intermediate $s$-channel consists of a string (or color
tube) with $q$ and $5q$ states at the ends.

It is interesting to compare the $u$-channel mechanism of the two-step
model described by Fig. 1 a) with the planar quark diagram of the QGSM
shown in Fig. 5. If the former desribes only one-nucleon exchange in
the $u$-channel, the latter is equivalent to an infinite sum of
contributions for all baryon resonances with isospin 1/2 lying on the
nucleon Regge trajectory.

\subsection{Spin structure of the $NN \rightarrow d a_0$
amplitude in the QGSM}

The spin dependence of the  $\gamma d \rightarrow pn$ amplitude
has been evaluated in Ref. \cite{GrishinaGDPN1}  by
assuming that all intermediate quark clusters have minimal spins
and the $s$-channel helicities in the quark-hadron and
hadron-quark transition amplitudes are conserved. In this limit
the spin structure of the amplitude $T(\gamma d \rightarrow pn)$
can be written as (see Ref. \cite{GrishinaGDPN1}, comment after
Eq.~(27))
\begin{eqnarray}
\langle p_3,\lambda_{p}; p_4,\lambda_{n} |
\hat{T}\left(s, t\right)|
p_2,\lambda_{d};p_1,\lambda_{\gamma}\rangle \simeq
\bar u_{\lambda_p}(p_3) \hat\epsilon_{\lambda_\gamma} \times\nonumber \\
\left[ A_{\gamma d \rightarrow pn}(s,t)( \hat{p}_3-\hat{p}_1)
 + B_{\gamma d \rightarrow pn }(s,t) m\right] \hat
{\epsilon}_{\lambda_{d}} v_{\lambda_n}(p_4), \label{spin1}
\end{eqnarray}
where $m$ is the nucleon mass, $p_1$, $p_2$, $p_3$, and $p_4$ are
the 4-momenta of the photon, deuteron, proton and neutron,
respectively, and $\lambda_i$ denotes the $s$ channel helicity of
the $i$-th particle. The invariant amplitudes $A_{\gamma d
\rightarrow pn}(s,t)$ and $B_{\gamma d \rightarrow pn}(s,t)$ have
similar Regge asymptotics (see below). It is possible to show (cf.
Ref. \cite{GrishinaGDPN1}) that at small scattering angles the
ratio $R_{\gamma d}=A_{\gamma d \rightarrow pn}(s,t)/B_{\gamma d
\rightarrow pn}(s,t)$ is a smooth function of $t$ and can be
considered as an effective constant that depends on the ratio of
the nucleon mass to the constituent quark mass $m_q$: $R\simeq
m/(2 m_q)$. We note that such a simple interpretation of $R$ in
general does not work at large scattering angles.

It is interesting to note that the spin structure of the  $\gamma
d \rightarrow pn$ amplitude in Eq.~(\ref{spin1}) is very similar
to the amplitude within the Reggeized Nucleon Born Term Approach
(RNBTA) where the $R_{\gamma d}=1$ is directly related to the spin
structure of the nucleon propagator (see Refs.
\cite{Guidal,Irving}).

In complete analogy with Eq.~(\ref{spin1}) the spin structure of
the amplitude $T(pp\rightarrow d a_0^+)$ can be written as
\begin{eqnarray}
\langle q_d,\lambda_d;q_{a_0} | \hat{T}\left(s,
t\right)| p_a,\lambda_a; p_b,\lambda_{b} \rangle \simeq
\bar v_{\lambda_a}(p_a) \hat{\epsilon}^{\,*}_{\lambda_d}
\times\nonumber \\
 \left[A_{pp\rightarrow da_0^+}(s,t)(\hat{p}_a - \hat{q}_{a_0}) + B_{pp\rightarrow da_0^+}(s,t) m\right] \hat{u}_{\lambda_b}(p_b)\ . \label{spina0}
\end{eqnarray}
In order to achieve consistency of the differential cross section
$d\sigma/dt$ with the Regge behaviour we use the following
parametrization of the amplitude $B_{pp\rightarrow d a_0^+}(s,t)$
\begin{equation}
\left|B_{pp\rightarrow d a_0^+}(s,t)\right|^{2}=\frac{1}{s} \
\left|\mathcal{M}_{\mathrm {Regge}}(s,t)\right|^2\ , \label{Bst}
\end{equation}
where
\begin{equation}
\mathcal{M}_{\mathrm{Regge}}(s,t)= F(t)
\left(\frac{s}{s_0}\right)^{\alpha_{N}(t)} \exp{\left[
      -i\ \frac{\pi}{2}\left(\alpha_{N}(t) -
        \frac{1}{2}\right)\right]}\  .
\label{eq:Mregge}
\end{equation}
Here $\alpha_N(t)$ is the trajectory of the nucleon Regge pole and
$s_0 =4~\mathrm{GeV}^2 \simeq m_d^2$.
We take the dependence of the residue $F(t)$ on $t$ in the form
\begin{equation}
  F(t) = B_{\mathrm{res}}
{\left[\frac{1}{m^2 - t}\ \exp{(R_1^2 t)} + C\, \exp{(R_2^2 t)} \right]}\
\label{eq:resid1}
\end{equation}
as used previously in Refs.~\cite{KaidalovP,Guaraldo} for the
description of the reactions $pp \rightarrow d \pi^+$ and $\bar p
d \rightarrow p \pi^-$ at $-t \leq 1.6$ GeV$^2$ as well as for the
analysis of deuteron photodisintegration at~$E_{\gamma} \ge 1$~GeV
(see Ref.~\cite{GrishinaGDPN1}). In Eq.~(\ref{eq:resid1}) the
first term in the square brackets contains the nucleon pole and
the second term accounts for the contribution of non-nucleonic
degrees of freedom in the deuteron.

The amplitudes defined by Eqs.~(\ref{spin1}) and (\ref{spina0})
have a rather simple covariant structure and can be extrapolated
to large angles. As shown in Ref. \cite{GrishinaGDPN1} the energy
behavior of the cross section for the reaction $\gamma d
\rightarrow pn$ at large angles crucially depends on the form of
the Regge trajectory $\alpha _N(t)$ for large negative $t$.
Best agreement with experimental data is obtained for a logarithmic 
form:
\begin{equation} \label{nonlin} \alpha_N(t)
= \alpha_N(0)- (\gamma \nu) \ln (1 - {t}/{T_B}) \ ,
\end{equation}
where the intercept $\alpha_{N}(0)=- 0.5$,
the slope $\alpha^{\prime}_{N}(0)=0.8 \div 0.9$~GeV${}^{-2}$ and
$T_B = 1.5 \div 1.7$~GeV${}^2$. We adopt the
following values for the parameters of the
residue $F(t)$ of Eq.~(\ref{eq:resid1}):
\begin{eqnarray}
&& C = 0.7\ \mathrm{GeV}^{-2}  , \;
R_1^2 = 1\div 2\ \mathrm{GeV}^{-2} , \;
R_2^2 = 0.03\  \mathrm{GeV}^{-2} \ . \nonumber
\end{eqnarray}
These parameters of the residue and trajectory, except for the
overall normalization factor $B_{\mathrm{res}}$, are not very
different from those determined by fitting data on the reactions
$pp \rightarrow  d \pi^+$ at $-t \leq 1.6\
\mathrm{GeV}^2$~\cite{KaidalovP} and $\gamma d\rightarrow pn$
at~$E_{\gamma}\ge 1$~GeV~\cite{GrishinaGDPN1}.

We considered the $pp\rightarrow d a_0^+$ amplitude (\ref{spina0})
within the RNBTA, i. e. for a fixed ratio $$R_{a_0
d}=A_{pp\rightarrow d a_0^+}(s,t)/B_{pp\rightarrow d
a_0^+}(s,t)=1,$$
as well as its generalization corresponding to the
QGSM. The spin structure of the amplitude within the QGSM takes
into account quark degrees of freedom and the parameter $R_{a_0
d}$ may be different from 1. In line with Ref. \cite{Grishina} we
also treat the ratio $R_{a_0 d}$ as a free parameter. The
parameters of the residue, trajectory and the ratio $R_{a_0 d}$
used for our calculations are given in Tables \ref{Tab1} and
\ref{Tab2}.

\begin{table*}[t]
\begin{center}
\begin{tabular}{|l|l|l|}
\hline Parameter & Set(${\gamma d}$) \cite{GrishinaGDPN1}&
Set(${a_0 d}$) \\ \hline $\alpha_{N}^{\prime}(0) \
[{\mbox{GeV}}^{-2} ]$ & 0.9 & 0.8 \\ $T_B \ [{\mbox{GeV}}^{2} ]$&
1.7 & 1.5 \\ $R_1^2 \ [{\mbox{GeV}}^{-2} ]$ & 2 & 1 \\ \hline
\end{tabular}
\end{center}
\caption{\label{Tab1} Parameters of the Regge trajectory
(\ref{nonlin}) and the residue (\ref{eq:resid1}) for the reactions
$\gamma d \rightarrow pn$ (Set($\gamma d$)) and $pp\rightarrow d
a_0^+$ (Set($a_0 d$)).}
\end{table*}

\begin{table*}[t]
\begin{center}
\begin{tabular}{|l|l|l|l|}
\hline
Parameters & RNBTM & \multicolumn{2}{|c|}{QGSM} \\
\hline
 trajectory \& residue &Set(${\gamma d}$) & Set(${\gamma d}$) & Set(${a_0 d}$)\\
$B_{{\mbox{res}}} \  [{\mbox{nb}}^{1/2}\cdot{\mbox{GeV}}^{3} ]$ & 5.23 $\times
10^{3}$
&3.19 $ \times 10^{3}$ &2.67$ \times 10^{3}$   \\
$R_{a_0 d}$ & 1 & - 4 & - 4 \\
\hline
\end{tabular}
\end{center}
\caption{\label{Tab2} Parameters of the trajectory and residue,
normalization factor $B_{{\mbox{res}}}$ and the ratio $R_{a_0 d}$
used for the $pp\rightarrow d a_0^+$ amplitude calculation within
the RNBTM and QGSM.}
\end{table*}

\subsection{Numerical results}

\begin{figure}[t]
  \centerline{\psfig{file=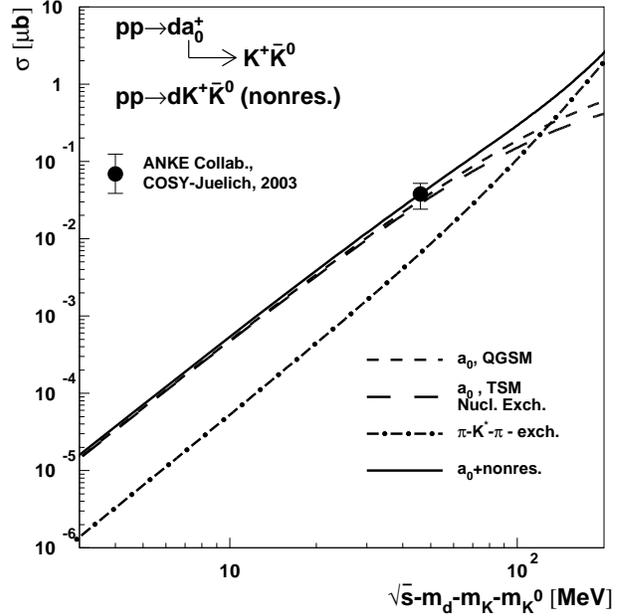,width=8cm}}
    \caption{Total cross section of the $pp \rightarrow d K^+ \bar{K}^0$
    reaction as a function of the c.m. excess energy. The long dashed
    line displays the $a_0$-resonance
    part of the cross section calculated within the TSM (same as in Fig.~(4)),
    which is very close to the results for the $a_0$ contribution
    from the  QGSM (short dashed line).
    The dash-dotted line shows the background
corresponding to the $\pi-K^{\star}-\pi$ exchange mechanism while
the solid line displays the sum of the background and the $a_0$
production cross section calculated within the QGSM. The full dot
shows the experimental cross section from Ref.
\protect\cite{Kleber}.}
  \label{fig:sigresnres}
\end{figure}

In Fig.~6 we show the $a_0$ resonance
contribution to the $pp \rightarrow d K^+ \bar{K}^0$ cross section
calculated within the  QGSM (dashed
curve) as well as the prediction of the TSM long-dashed
line). The dash-dotted line displays the background corresponding
to the $\pi-K^{\star}-\pi$ exchange mechanism. Since we have $K
\bar {K}$ pairs in a relative S-wave basically due to direct $a_0$
resonance production, we have normalized the results of the 
QGSM at $Q=46$~MeV to the experimental value $31.5$~nb, which
was found for the $K\bar{K}$ S-wave part  \cite{Kleber}. The
corresponding values of the normalization factor $B_{\mbox{res}}$
are given in Table \ref{Tab2}. In Fig.~6   we
display the result of the QGSM with parameters of Set ($a_0
d$). Since the calculations with Set($\gamma d$) give practically
the same answer we discard an explicit representation in this
figure. As seen from Fig.~6 the energy
dependence of the $a_0$ resonance contribution of the cross
section predicted by the TSM and QGSM is very similar at $Q\leq 200$
MeV. The solid line in Fig.~ 6 displays the sum of the $a_0$
resonance production cross section calculated within the QGSM and
the $K\bar K$ $P$-wave background contribution.

\begin{figure}[t]
  \centerline{\psfig{file=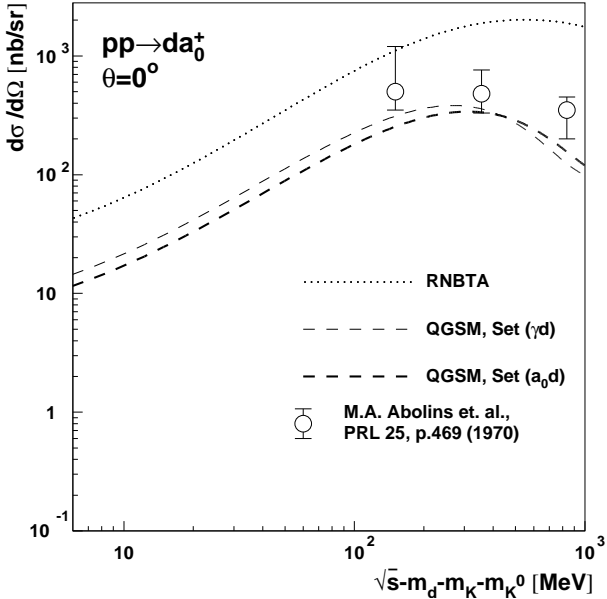,width=8cm}} 
  \caption{Forward differential cross section of the reaction
  $pp\rightarrow d a_0^+$ as a function of the c.m.excess energy.
  The open dots are the experimental data from
  Ref.~\protect\cite{Abolins}. The dotted line shows the prediction of
  the RNBTA. The thin and bold dashed curves display the results of
  the QGSM with parameters of  Set($\gamma d$) and Set($a_0d$),
  respectively.}
\label{fig:dsacosy}
\end{figure}

In order to check the consistency of our model for the $a_0$
production in the $pp\rightarrow d a_0^+$ reaction we compare the
calculated forward differential cross section with the LBL
data \cite{Abolins} in Fig.~\ref{fig:dsacosy}. The dotted line
shows the prediction of the RNBTA. The calculations within the
QGSM --- normalized to the ANKE data on the reaction $pp\rightarrow
d a_0^+ \rightarrow K^+ \bar K^0$ --- are in a good agreement with
the differential cross sections measured at LBL \cite{Abolins}
(open circles).

\begin{figure}[t]
  \centerline{\psfig{file=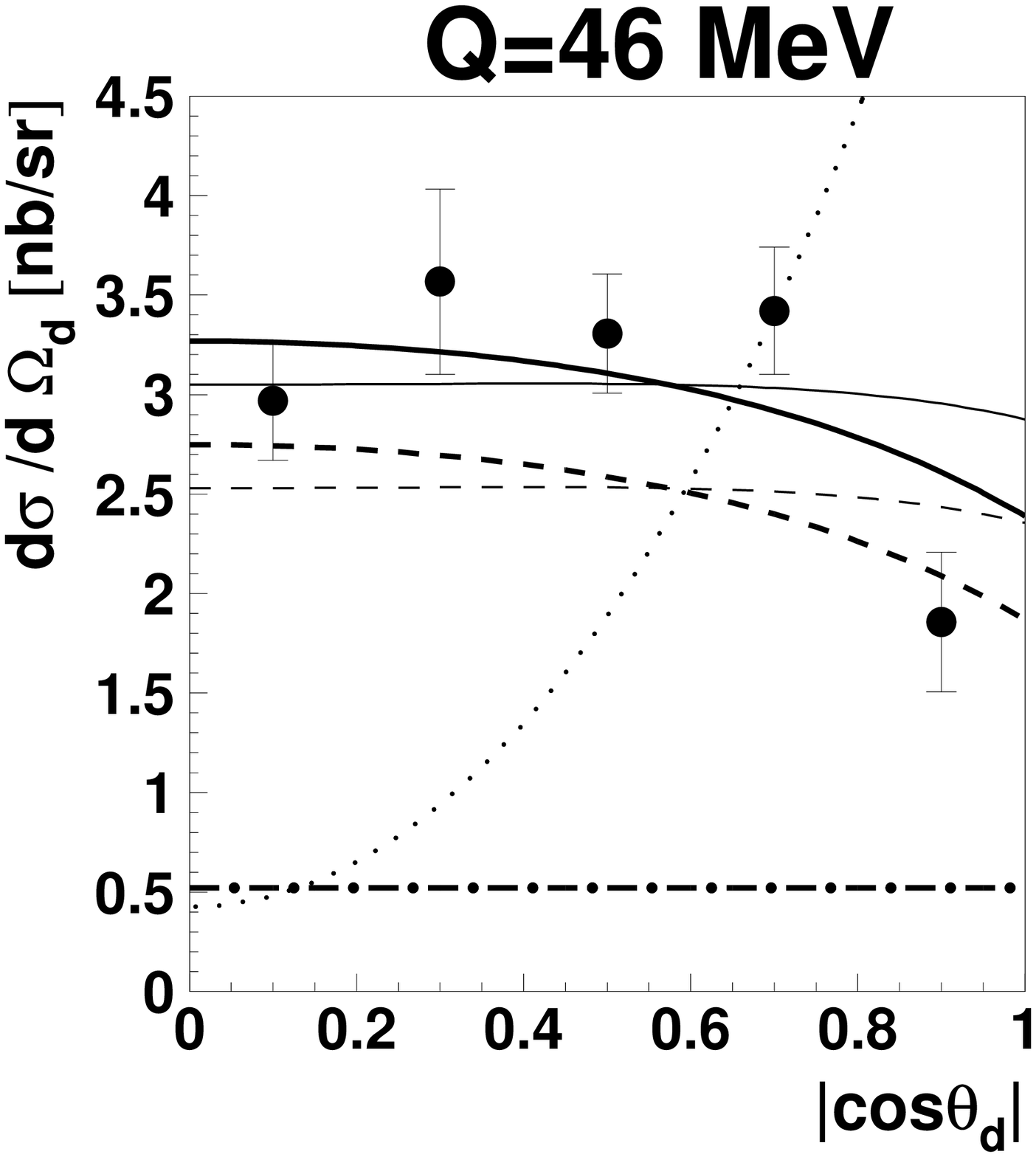,width=5.0cm}}
  \centerline{\psfig{file=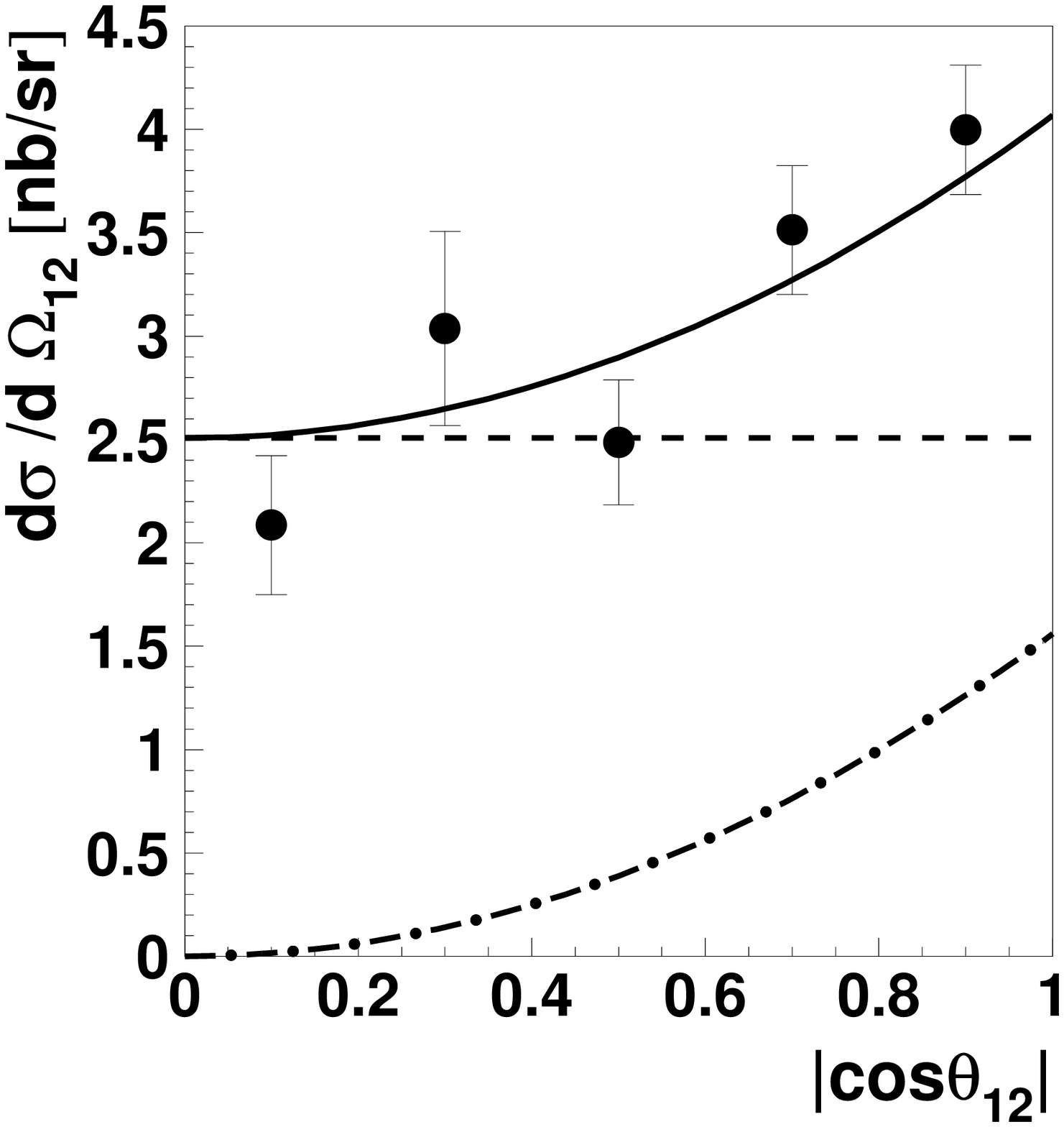,width=5.0cm}}
  \centerline{\psfig{file=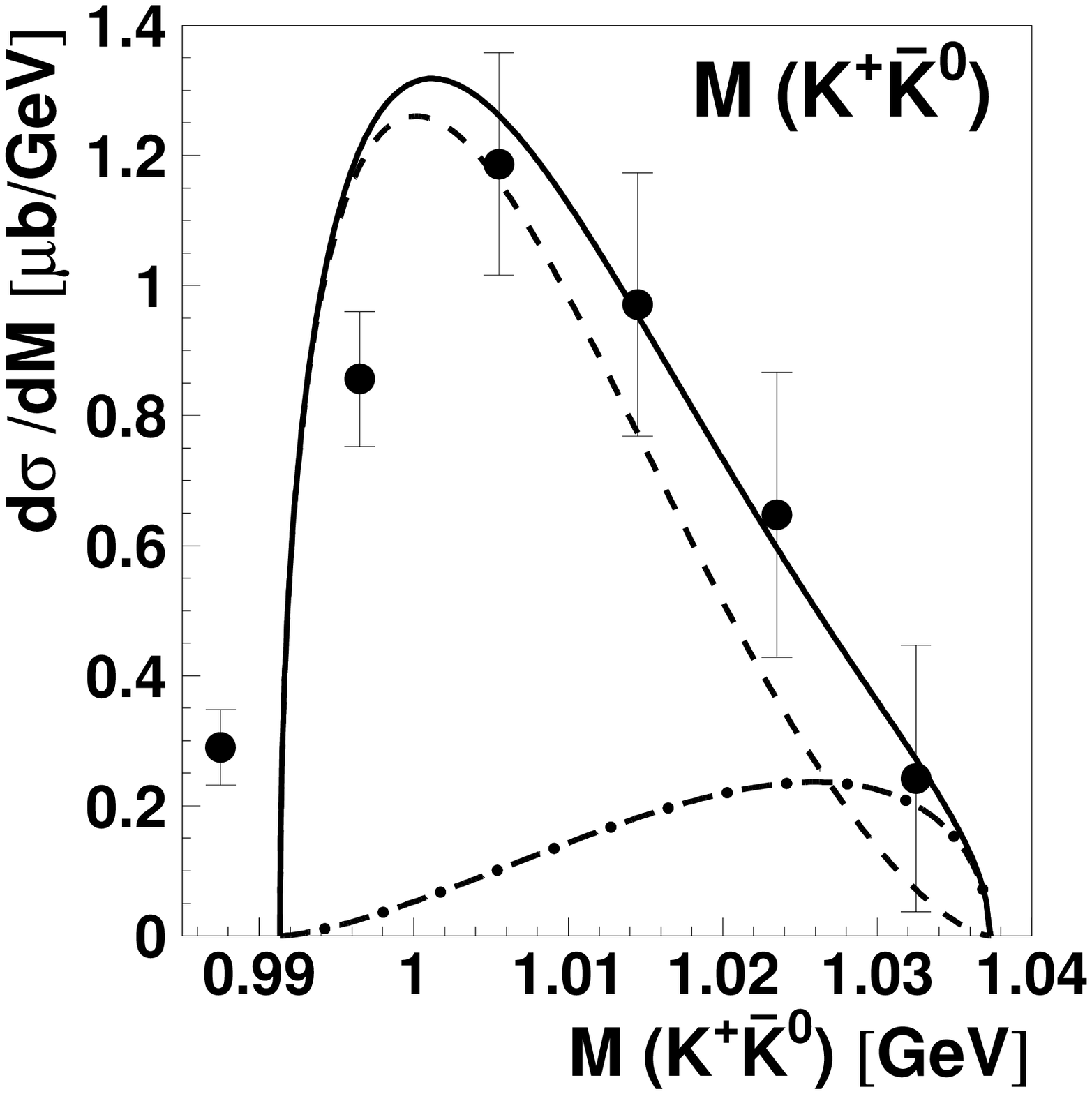,width=5.0cm}}
  \caption{Angular distributions (upper and middle part) and invariant
  mass distribution (lower part) for the $pp \rightarrow d K^+
  \bar{K}^0$ reaction at $Q=46$ MeV in comparison with the data from
  Ref.~\cite{Kleber}.  The dashed (dashed-dotted) line corresponds to
  $K^+ \bar{K^0}$ production in a relative $S$-($P$-) wave and the solid
  line is the sum of both contributions. The $a_0$-resonance
  contribution shown by the bold and thin dashed lines results from
  the QGSM with parameters of Set($a_0d$) and Set($\gamma d$),
  respectively. The dotted line is the result from the RNBTA.
  $\Theta_d$ and $\Theta_{12}$ are the polar angles for the
  c.m.\ deuteron momentum and for the $K \bar K$ relative momentum, 
  respectively.}
 \label{fig:distr46QGSM}
\end{figure}

The calculated angular and invariant mass distributions for the
$pp \rightarrow d K^+ \bar{K}^0$ reaction at $Q=46$ MeV in
comparison to the experimental data \cite{Kleber} are shown in
Fig. \ref{fig:distr46QGSM}. The dashed lines correspond to $K^+
\bar{K}^0$ production through the $a_0$ resonance and has been
calculated within the QGSM using the parameters from Set($a_0
d$). The dashed-dotted lines describe the $K \bar K$ $P$-wave
background calculated within the $\pi-K^{\star}-\pi$-exchange
model. The solid lines indicate the sum of the $a_0$ resonance and
background contributions. In the upper part of the figure we show
also the angular distribution for deuterons calculated in the QGSM
with parameters of Set($\gamma d$). The almost isotropic
angular dependence given by this version of the QGSM (thin solid
line) is in a reasonable agreement with the data. The angular
distribution of deuterons for the $a_0$ contribution as calculated
within the RNBTA is presented by the dotted line and gives a sharp
forward peak similarly to the nonrelativistic two-step model
\cite{Grishina}. Therefore, both models --- TSM and RNBTA --- are
not able to reproduce the experimental deuteron angular
distribution \cite{Kleber}. The best description of the data (bold
solid line) is obtained  by the QGSM with parameters of the
Set~($a_0 d$).

Therefore, the QGSM gives a rather good description of the ANKE
data on the reaction $pp \rightarrow dK^+\bar{K}^0$  at $Q$=46 MeV
\cite{Kleber} simultaneously in agreement with the forward differential cross
section of the reaction $pp\rightarrow da_0^+$  measured at
LBL at 3.8, 4.5 and 6.3~GeV/c \cite{Abolins}.

\begin{figure}[t]
  \centerline{\psfig{file=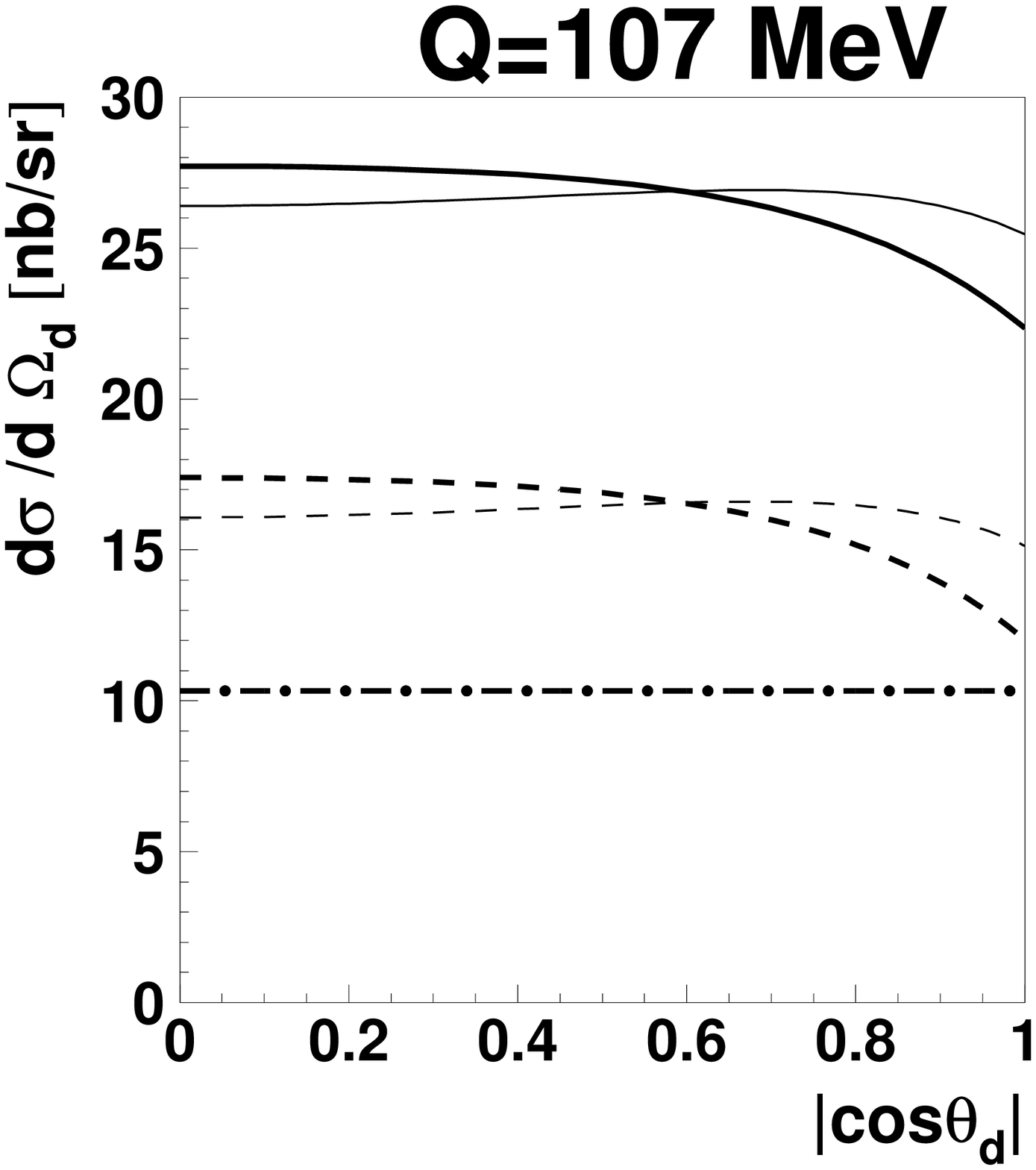,width=5.0cm}}
  \centerline{\psfig{file=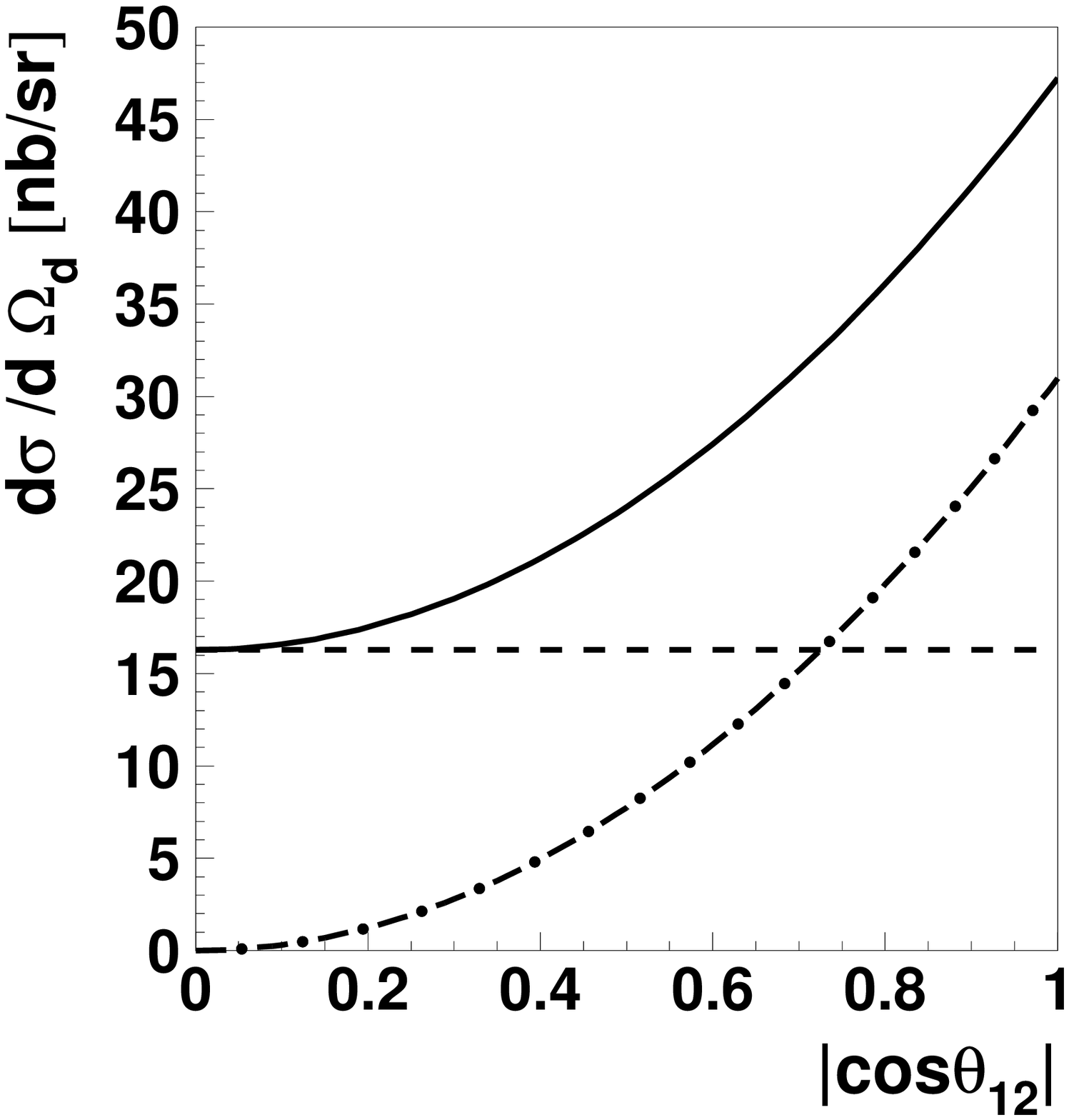,width=5.0cm}}
  \centerline{\psfig{file=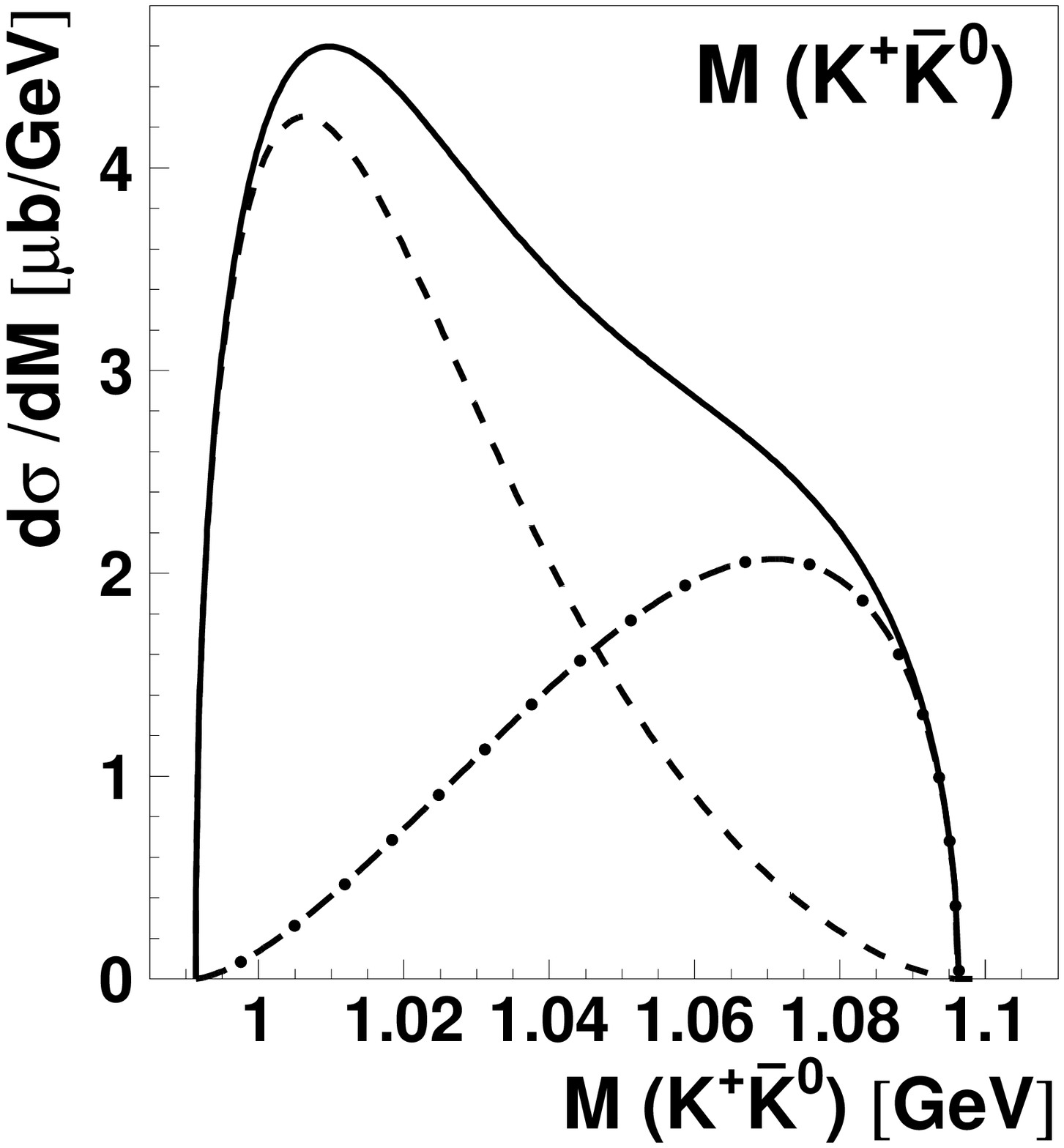,width=5.0cm}}
    \caption{Angular distributions (upper and middle part) and
    invariant mass distribution (lower part)
    for the $pp \rightarrow d K^+ \bar{K}^0$
    reaction at $Q=107$ MeV (see  Fig.~\ref{fig:distr46QGSM} for
    the description of the lines.)}
  \label{fig:distr110QGSM}
\end{figure}

In Fig.~\ref{fig:distr110QGSM} we present the predictions for the
angular and mass distributions at $Q=107$ MeV, where corresponding
experimental data from ANKE are expected soon. It is important to
note that our model for the $pp\rightarrow d \bar {K}^0 K^+$
reaction predicts that the ratio of the background  to the $a_0$
contribution will increases by a factor of 3. Therefore, the
background contribution is expected to be about 40 \% at
$Q=107$~MeV. As seen from the lower part of
Fig.~\ref{fig:distr110QGSM} the $a_0$ resonance part can be
separated from the contribution from the $K^+ \bar{K}^0$ $P$-wave
background: Most of the events related to the $a_0$ resonance are
concentrated in the lower part of the $K^+ \bar{K}^0$ mass
spectrum, whereas the main contribution of the background shows up
at higher invariant mass.

\section{Final state interactions}

As has been stressed in Ref. \cite{Oset} the reaction $pp \to d K^+
\bar{K^0}$ might be sensitive to both the  $K^+ \bar{K^0}$ and
$\bar K d$ final-state interactions (FSI). The interaction of the
$K^+$ with protons and neutrons is rather weak \cite{Kaiser} and
following Ref.~\cite{Oset} we will neglect it. Within our model we can
describe the $S$-wave $K \bar K$ cross section by direct $a_0^+$
production with subsequent decay $a_0^+ \rightarrow K^+
\bar{K^0}$. Contributions from non-resonant $S$-wave $K \bar K$
production turned out to be negligeably small, whereas the $P$-wave $K
\bar K$ FSI it is small due to centrifugal suppression. Thus we only
have to consider the $\bar K d$ FSI. To estimate the role of the
$S$-wave $\bar K d$ FSI we use the Foldy-Brueckner adiabatic approach
based on the multiple scattering (MS) formalism (see
Ref.~\cite{Goldberger}).  Note that this method has already been used
for the calculation of the enhancement factor for the reactions $p d
\to \mathrm{^{3}He}\eta$~\cite{Faldt2} and $pn\to d\eta$~\cite{Grishina1}.

In the Foldy-Brueckner adiabatic approach the $\bar{K}^0 d $ wave
function --- defined at fixed coordinates of the proton
($\mathbf{r}_p$) and the neutron ($\mathbf{r}_n$) (see
Ref.~\cite{Goldberger} for details) --- reads as:
\begin{eqnarray}
&&  \Psi_k(\mathbf{r}_{\bar{K}^0},\mathbf{r}_p,
\mathbf{r}_n) = \exp{(i  \mathbf{k}  \mathbf{r}_{\bar{K}^0})}
    + \frac{t_{\bar{K}^0 p}}{D}\, \frac{\exp{(ik r_{\bar{K}^0
p})}}{r_{\bar{K}^0 p}} \,  \nonumber \\
&\times&
 \left( \exp{(i\mathbf{k} \mathbf{r}_p)} + t_{\bar{K}^0 n}\,
  \frac{\exp{(ikr_{pn})}}{r_{pn}}\exp{(i\mathbf{k} \mathbf{r}_n)} \right) \nonumber \\
 &+& \frac{t_{\bar{K}^0 n}}{D}\, \frac{\exp{(ikr_{\bar{K}^0 n})}}{r_{\bar{K}^0 n}}
 \nonumber \\    &\times&  \left(
  \exp{(i\mathbf{k} \mathbf{r}_n)} + t_{\bar{K}^0 p} \frac{\exp{(ikr_{pn})}}{r_{pn}}
 \exp{(i\mathbf{k} \mathbf{r}_p)} \right)
 \ ,
\end{eqnarray}
where
\begin{equation}
  D = \left( 1 - t_{\bar{K}^0 p}t_{\bar{K}^0 n}\,
    \frac{\exp{(2ikr_{pn})}}{r^2_{pn}} \right) \ .
\end{equation}
Here $\mathbf{r}_{pn}=\mathbf{r}_p - \mathbf{r}_n$,
$\mathbf{r}_{\bar{K}^0 p}=\mathbf{r}_{\bar{K}^0} - \mathbf{r}_p$,
$\mathbf{r}_{\bar{K}^0 n}=\mathbf{r}_{\bar{K}^0} - \mathbf{r}_n$
and $\mathbf{k} = \mathbf{q}_{1d} \,
\frac{m_d+m_{\bar{K}^0}}{m_d}$ and $k$, $r_{pn}$, etc., are the
moduli of these vectors; $t_{\bar{K}^0 N}$ is the $\bar{K}^0 N$
$t$-matrix which is related to the scattering amplitude
$f_{\bar{K}^0 N}$ by \begin{equation}
t_{\bar{K}^0 N}(k_{\bar{K}^0 N}) =
(1+\frac{m_{\bar{K}^0}}{m})\, f_{\bar{K}^0 N}(k_{\bar{K}^0 N}).
\end{equation}
Note that we use the unitarized scattering length approximation
for the latter, i.e. \begin{equation}
 f^{I}_{\bar{K} N}(k_{\bar{K} N})=
\left((a^{I}_{\bar{K} N})^{-1} - ik_{\bar{K} N}\right)^{-1},
\end{equation}
 where
$k_{\bar{K} N}$ is the modulus of the relative $\bar{K} N$
momentum and $I$ denotes the isospin of the $\bar{K} N$ system.

The  $\bar{K}^0 d$-scattering length then is defined as
\begin{eqnarray} \label{FSI0}
 && A^{\mathrm{MS}}_{\bar{K}^0 d} = \frac{m_d}{m_{\bar{K}^0}+m_d}  \nonumber \\
&& \times \left<\frac{\displaystyle t_{\bar{K}^0 p}(k_{\bar{K}^0
p} =0)+ t_{\bar{K}^0 n}(k_{\bar{K}^0 n} =0)+ t_r } {\displaystyle
1- {t_{\bar{K}^0 p}(k_{\bar{K}^0 p} =0) t_{\bar{K}^0
n}(k_{\bar{K}^0 n} =0)}/{r^2} }
  \right>  \ ,
\end{eqnarray}
and the FSI enhancement factor as
  \begin{eqnarray}
\lambda^{\mathrm{MS}}(q_{1d}) =
  | \langle \Psi_{k}(\mathbf{r}_{\bar{K}^0}=0,\mathbf{r}_p=
  \mathbf{r}/2, \mathbf{r}_n =- \mathbf{r}/2) \rangle |^2 \ .
\end{eqnarray}
In Eq. (\ref{FSI0}) we have used the abbreviation
\begin{equation}
t_r= \frac{2 t_{\bar{K}^0
p}(k_{\bar{K}^0 p} =0) t_{\bar{K}^0 n}(k_{\bar{K}^0 n} =0)}{r}\ .
\end{equation}
To describe the deuteron structure we use the Paris wave function
\cite{Lacomb}. The $\bar{K}N$ scattering lengths $a^0_{\bar{K}N}$ and
$a^1_{\bar{K}N}$ are taken from Ref.~\cite{Deloff}:\\ i) $a_0
=-1.57~+i~0.78~ \mathrm{fm}$, $a_1 =0.32~+i~0.75~ \mathrm{fm}$ (CSL
set);\\ ii)) $a_0 =-1.59~+i~0.76~ \mathrm{fm}$, $a_1 =0.26~+i~0.57~
\mathrm{fm}$ ( $K$-matrix set).

We recall that the $\bar{K}N$ scattering length is strongly repulsive
for the isospin channel $I$=0 and moderately attractive for $I$=1. In
the single scattering approximation then a slight repulsion adds up
for the $\bar{K}d$ system $ A_{\bar K d}^{IA}= -0.39 + i~1.72~
\mathrm{fm}$~\cite{Deloff}. Results from Faddeev calculations with
separable $\bar K N$ potentials --- as carried out in
Ref. \cite{Torres} --- give $ A_{\bar K d}= -1.34 + i~1.04~
\mathrm{fm}$, i.e., they predict a larger $\bar K d $
repulsion. We remind the reader that a repulsion in the low-energy
$\bar K d$ system can lead to a FSI suppression factor ($< 1$); on the
other hand, any attraction leads to a FSI enhancement factor ($> 1$).

Evidently, the FSI effect is most important close to threshold
and is due to the long-range coherent $S$-wave $\bar K d$
interaction. Therefore, one can safely assume that the range of
the FSI is much larger than the range of the 'hard' interaction,
which is responsible for the production of the $K \bar K$-meson
pair. In this case the basic production amplitude and the FSI can
be factorized \cite{Goldberger}, i.e. the FSI can be taken into
account by multiplying the production cross section by the FSI
factor.

The partial wave structure of the final state for the basic production
amplitude corresponds to $[(\bar{K^0}K^+)_s d]_P$, for $a_0$
production and $[(\bar{K^0}K^+)_p d]_S$ for the $K \bar K$
background. To calculate the corresponding FSI factors we expressed
these partial waves in terms of partial amplitudes of the second basis
with $[(d\bar{K^0})_sK^+]_P$ and $[(d\bar{K^0})_pK^+]_S$.  Then we
have to take into account that only the first term of the second basis
is renormalized due to the $S$-wave $\bar K d$ interaction (see
e.g. Ref.~\cite{Oset}).  According to experimental data \cite{Kleber}
the latter configuration gives about 50\% contribution to the total
production cross section of the reaction $pp \to d K^+ \bar{K^0}$ at
$Q$=46 MeV~\cite{Hanhart}.
\begin{figure}[t]
  \centerline{\psfig{file=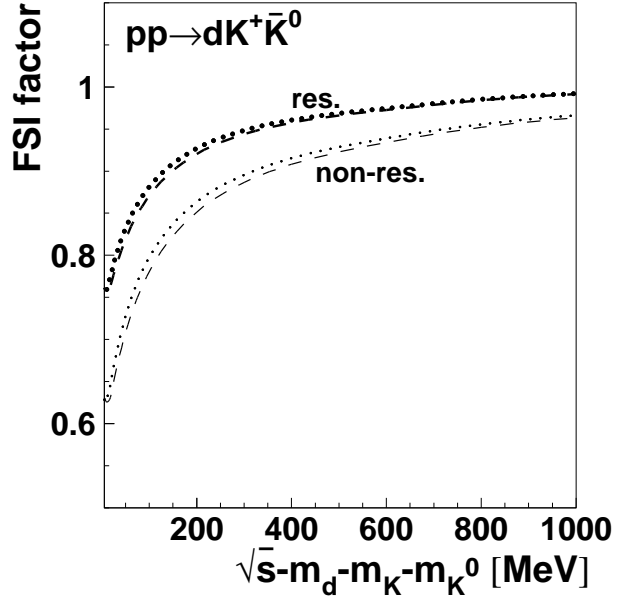,width=8cm}}
    \caption{The final state interaction factor for the reaction $pp
    \to d K^+ \bar{K^0}$ as a function of the energy above
    threshold. The upper and lower lines correspond to $a_0$
    production and the $K \bar K$ background, respectively. The dashed
    and dotted lines correspond to the CSL and K-matrix sets of the
    $\bar K N$ scattering length \cite{Deloff}, respectively.}
 \label{fig9}
\end{figure}

\begin{figure}[t]
  \centerline{\psfig{file=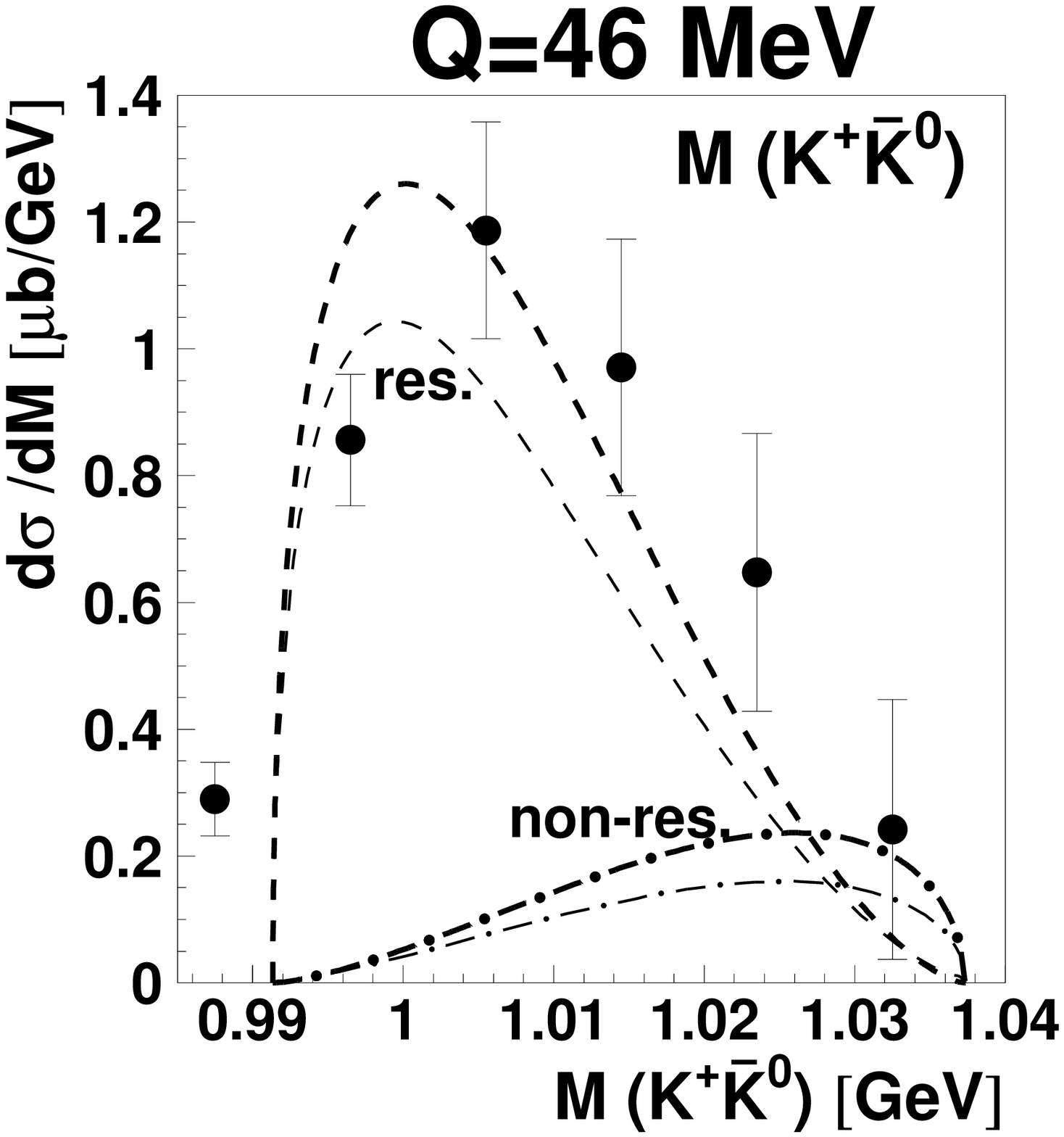,width=7cm}}
  \centerline{\psfig{file=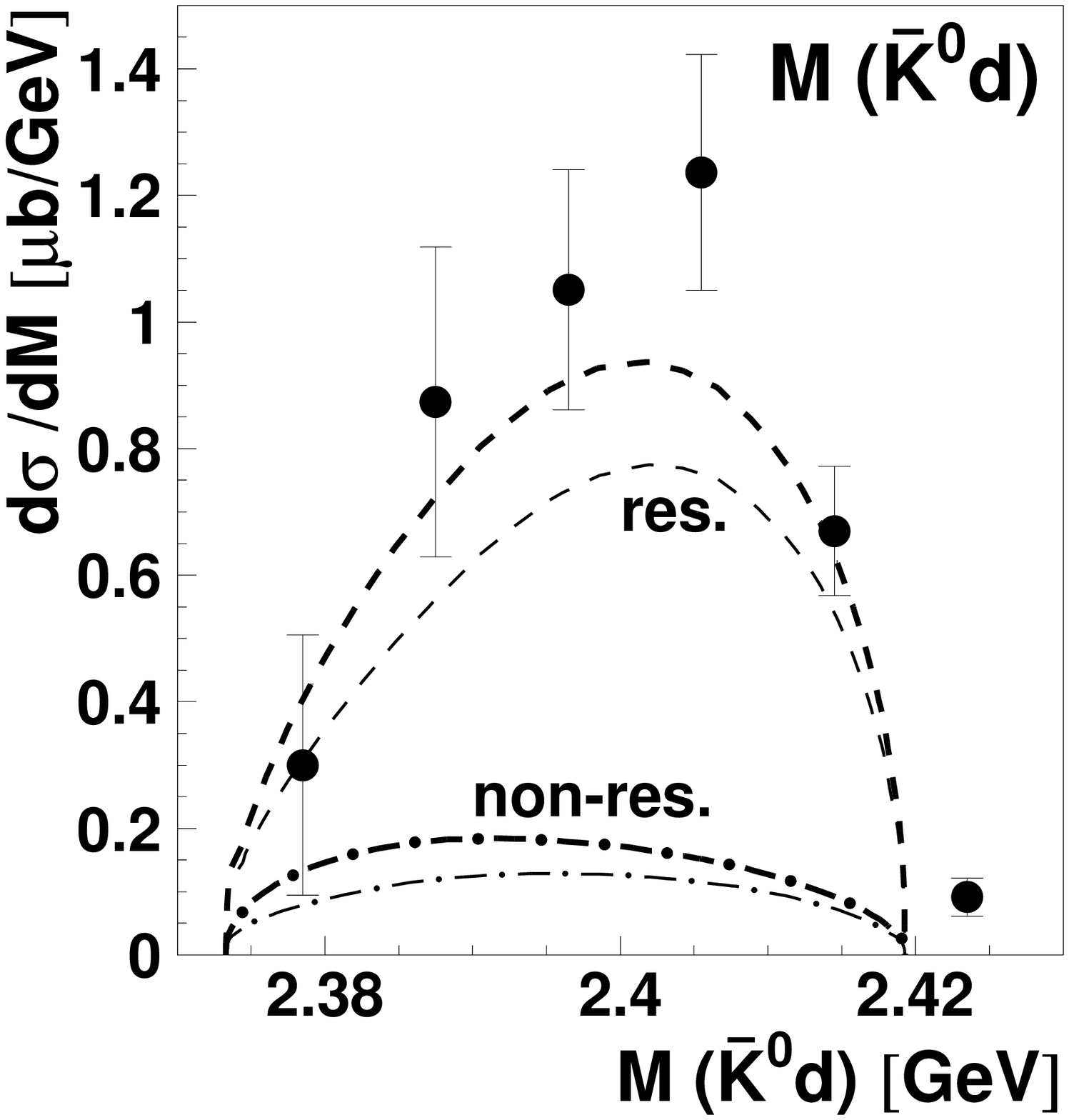,width=6.6cm}}
  \caption{Invariant mass distributions for the $K^+ \bar{K^0}$
  (upper part) and $d \bar{K^0}$ (lower part) systems for the reaction
  $pp \to d K^+ \bar{K^0}$ at $Q$=46 MeV. The dashed (dash-dotted)
  lines are calculated for the resonance (non-resonance)
  contributions. The bold (thin) curves describe the contributions
  without (with) the FSI included. The experimental data are taken
  from Ref.~\protect\cite{Kleber}.}
  \label{fig10}
\end{figure}

\begin{figure}[t]
  \centerline{\psfig{file=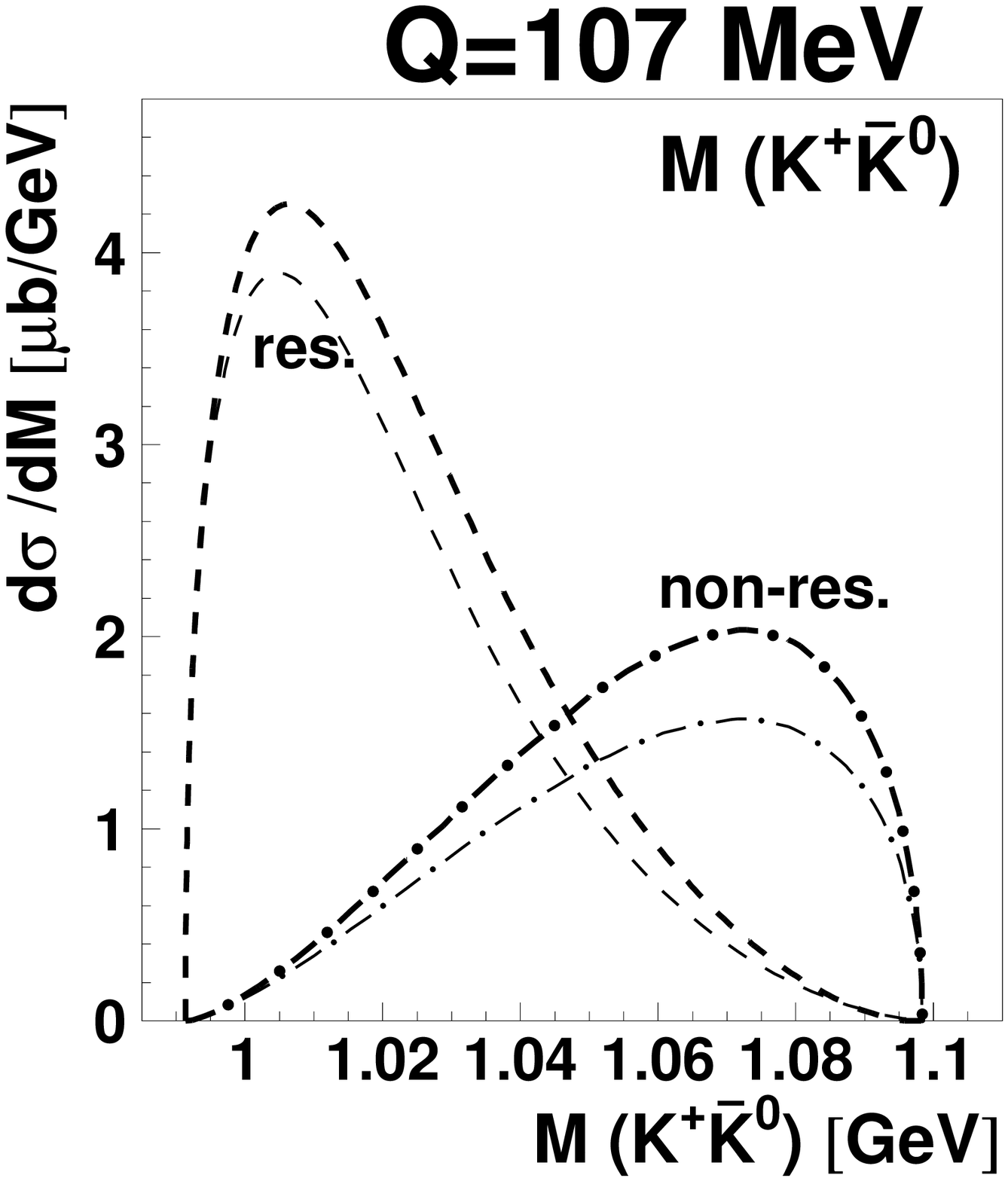,width=6cm}}
  \centerline{\psfig{file=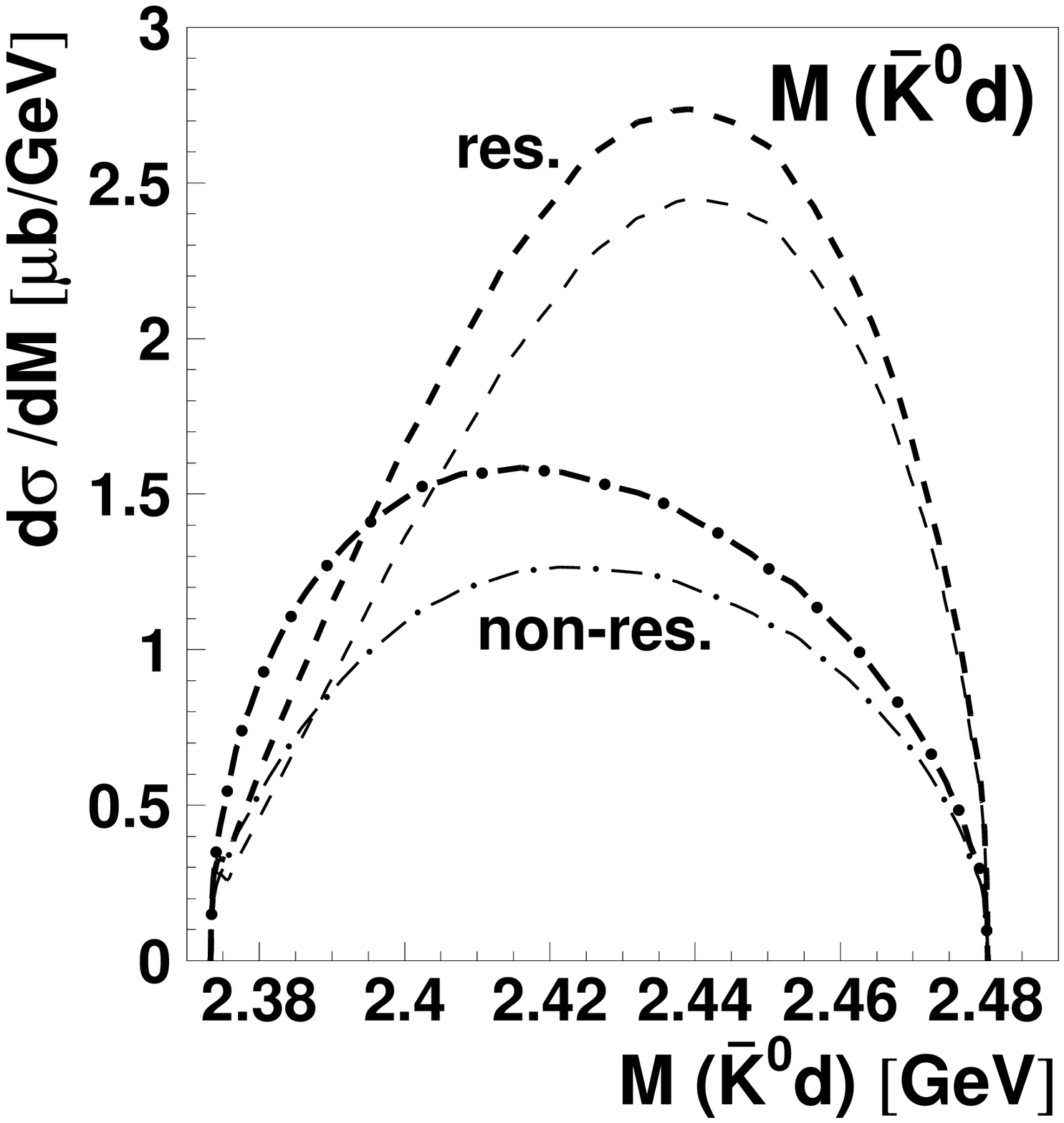,width=6cm}}
    \caption{Invariant mass distributions for the $K^+ \bar{K^0}$
    (upper part) and $d \bar{K^0}$ (lower part) systems for the
    reaction $pp \to d K^+ \bar{K^0}$ at $Q$=107 MeV. The assignment
    of the individual lines is the same as in Fig.~\ref{fig10}.}
    \label{fig11}
\end{figure}

The results of our calculations for the FSI effect on the cross
section of the reaction $pp \to d K^+ \bar{K^0}$ as well as on the
$K^+ \bar{K^0}$ and $d \bar{K^0}$ mass distributions are shown in
Figs.~\ref{fig9}, ~\ref{fig10} and \ref{fig11}. We start with the
energy dependence of the FSI factor which is presented in
Fig.~\ref{fig9}. The upper and lower lines correspond to $a_0$
production and the $K \bar K$ background, respectively. We find that
the FSI factors are smaller than one as expected from the repulsion in
the system (see discussion above). Furthermore, the suppression of the
non-resonant background is larger than for the $a_0$ resonant
channel. In the latter case the suppression is about 0.81 at $Q$=46
MeV and 0.88 at 107 MeV, while the background is suppressed by 0.7 at
$Q$=46 MeV and 0.79 at 107 MeV, respectively. The dashed and dotted
lines correspond to the CSL and K-matrix sets of the $\bar K N$
scattering length \cite{Deloff}; both parameter-sets lead to
approximately the same suppression factors.

The invariant mass distributions for the $K^+ \bar{K^0}$ and  $d
\bar{K^0}$ systems are shown in Figs.~\ref{fig10} and \ref{fig11}
for $Q$=46 MeV and 107 MeV, respectively. The dashed (upper) lines
are calculated for the resonance  contributions, while the
dash-dotted (lower) lines stand for the non-resonance
contributions. The bold lines describe the contributions
calculated without FSI, where the (thin) lines with FSI are
always slightly lower in line with Fig.~\ref{fig9}. 

We note that the QGSM cannot predict the absolute value of the cross
section and has been 'normalized' to the data at 46 MeV. If we rescale
the respective mass distribution up by $\sim$ 20 \% we obtain
distributions practically identical to the bold dashed lines
calculated without FSI. Therefore, increasing the normalization of the
QGSM by 1.2 our calculations for the $K^+ \bar{K^0}$ and $d
\bar{K^0}$ mass distributions will be again in a good agreement with the
ANKE data \cite{Kleber}. Let us note that the predictions of
Ref.~\cite{Oset} on strong distorsions of the $K^+ \bar{K^0}$
and $\bar{K^0} d$ invariant mass spectra by the $\bar{K^0} d$ FSI were
not confirmed by the experiment~\cite{Kleber}.

We finally address the validity of the FSI model employed here.
The multiple scattering (or fixed center) approach (MSA) was
applied to the calculations of the $K^- d$ scattering length in
Ref.~\cite{Deloff} before and has also been compared to full
multichannel Faddeev calculations in Ref.~\cite{Deloff2}. In the
latter studies it was found that the MSA --- with a single-channel
absorptive $\bar K N$ interaction --- gives quite reliable
estimates for the real and imaginary parts of the $K^- d$
scattering length. Our results for the latter are in reasonable
agreement with the calculation of Ref. \cite{Deloff}: we found $
A_{\bar K d}= - 0.78 + i~1.23~ \mathrm{fm}$ for the $K$-matrix set
while  Ref. \cite{Deloff} gives $ A_{\bar K d}= -0.72 + i~0.94~
\mathrm{fm}$ which has to be multiplied additionally by the
'reduced mass' factor (see, e.g., \cite{Faldt2})
\begin{equation}
\frac{(1+ m_{\bar{K^0}} /m)}{(1+ m_{\bar{K^0}} /m_d)} \simeq 1.18.
\end{equation}
This gives  $ A_{\bar K d}^* = -0.85 + i~1.11~ \mathrm{fm}$. The
agreement with our result is evidently quite good.
 
\section{Conclusions}

In this work we have performed a detailed study of $a_0$
production in the reaction $NN \rightarrow  d K^+\bar K^0 $ near
threshold and at medium energies. Using the two-step model (TSM)
based on an effective Lagrangian approach with one-pion exchange
in the intermediate state we have analyzed different contributions
to the cross section of the reaction $NN\rightarrow d a_0$
corresponding to $t$-channel diagrams with $\eta$- and
$f_1(1285)$-meson exchanges as well as $s$ and $u$-channel graphs
with an intermediate nucleon. We have also considered  the
$t$-channel Reggeon mechanism with $b_1$ and $\rho_2$ exchanges
with parameters normalized to the Brookhaven data for
$\pi^-p\rightarrow a_0^0n$ at 18 GeV/c \cite{Dzierba}. These
results have been used to calculate the contribution of $a_0$
mesons to the cross section of the reaction $pp\rightarrow d
K^+\bar K^0$. We found that the dominant contribution is given by
the nucleon $u$-channel mechanism.

Within this approach, which is
practically equivalent to a direct normalization of the
$u$-channel contribution to the LBL data~\cite{Abolins} on
the forward differential cross section of the reaction $pp
\rightarrow d a_0^+$ at 3.8 GeV/c,  we could reproduce fairly well the
total cross section of the reaction  $pp\rightarrow d K^+\bar K^0$
at 3.46 GeV/c ($Q$ = 46 MeV) as measured at COSY \cite{Kleber}.
However, the TSM failed to reproduce the experimental distribution
in the deuteron scattering angle.

As an alternative and more general approach we have employed the
Quark-Gluon Strings Model (QGSM), that recently has successfully been
applied to the description of deuteron photodisintegration data
\cite{GrishinaGDPN1,GrishinaGDPN2}. Within the QGSM there is an almost
complete analogy between the amplitudes of the reactions $\gamma d
\rightarrow pn$ and $NN \rightarrow d a_0$ because both are described
by planar graphs with three valence-quark exchange in the $t$ (or
$u$)-channels (cf. Fig. 5). Normalizing the QGSM predictions to the
total cross section of the reaction $pp \rightarrow d a_0^+
\rightarrow d K^+\bar K^0 $ at $Q$ = 46 MeV we have calculated the
energy dependence of the cross section as well as the angular and mass
distributions at $Q$= 46 and 107 MeV.  In the QGSM we were able to
reproduce the differential experimental distributions at $Q$=46
MeV. We have, furthermore, demonstrated that the QGSM gives also a
rather good description of the LBL data at intermediate energies. In
order to test the QGSM and its implications we have made detailed
predictions for an excess energy of 107 MeV that can be controlled
experimentally in the near future.

We also analyzed the non-resonant $K \bar{K}$-pair production
using a model with $\pi-K^{\star}-\pi(\eta)$- and $K$-exchange
mechanisms. It is found that the $K$-exchange mechanism can be
neglected. As following from
$G$-parity conservation arguments
the $\pi-K^{\star}-\pi$ mechanism contributes mainly to the $P$-
wave in the $K^+ \bar{K}^0$-system, while the
$\pi-K^{\star}-\eta$-mechanism contributes dominantly to the $S$-wave.
The latter channel turned out to be negligibly small.
In addition we have explored the effects from final-state
interactions (FSI) in these reactions for the resonant and
non-resonant channels. Due to an effective repulsive interaction
in the $\bar{K} d$ system the FSI factor
turns out to be smaller than one. However, the net suppression found
is only in the order of 20\% for the $a_0$ channel, while the
background is suppressed by up to $\sim$ 30\%. Moreover, the shape 
of the invariant mass distributions in the $K^+ \bar{K^0}$ and 
$\bar{K^0} d$ channels is practically not influenced by the FSI.

In summary, we conclude that the reaction $pp \rightarrow d K^+
\bar{K}^0$ at excess energies $Q \leq 100$ MeV should be dominated by
the intermediate production of the $a_0(980)$-resonance. For $Q \geq
100$~MeV the non-resonant $K^+ \bar{K}^0$-pair production can be
important, however, this background gives a dominant contribution to
the $K^+
\bar{K}^0$ $P$-wave at higher $K^+ \bar {K}^0$ invariant mass.
This implies that the experimental program on the study of
near-threshold $a_0$ and $f_0$ production in $pp,~pn,~pd$ and $dd$
interactions at COSY-J\"ulich \cite{Buescher,Buescher1} is
promising since the $a_0$ signal in the $K \bar K$ mode can
reliably be separated from the non-resonant $K \bar{K}$
background.

\section*{Appendix}
In this appendix we present the $\pi N \rightarrow N a_0$
amplitudes which were used in Section~3 for the calculation of the
resonant contribution to the reaction $pp  \rightarrow  dK^+
\bar{K^0}$.

The $t$-channel $f_1(1285)$ and $\eta$ exchanges are
described by the expressions
\begin{eqnarray}
&&\hspace*{-4mm}M_\eta^t(\pi^-p\rightarrow a_0^- p) = g_{\eta\pi
a_0} g_{\eta NN}\ \bar u(p_2^\prime) \gamma_5 u(p_2)
\nonumber\\ &\times& {1\over t-m_\eta ^2} \ F_{\eta\pi a_0}(t)
F_{\eta NN}(t), \label{eq2}
\end{eqnarray}
\begin{eqnarray}
&&\hspace*{-4mm}M_{f_1}^t(\pi^- p\rightarrow a_0^- p) = g_{f_1\pi
a_0} g_{f_1NN}   \nonumber\\ &\times& (p_1+p_1^\prime)_\mu \
\left(g_{\mu\nu}-{q_\mu q_\nu\over m_{f_1}^2}\right) \ \bar
u(p_2^\prime) \gamma_\nu \gamma_5 u(p_2)
        \nonumber\\
&\times& {1\over t-m_{f_1}^2}\ F_{f_1\pi a_0}(t) \ F_{f_1NN}(t).
\label{eq3}
\end{eqnarray}
Here $p_1$ and $p_1^\prime$ are the four momenta of $\pi^-$ and
$a_0^-$, whereas $p_2$ and $p_2^\prime$ are the four momenta of the
initial and final protons, respectively, and $q=p_2^\prime-p_2$,
$t=(p_2^\prime-p_2)^2$. The form factors $F_j(t)$ at the different
vertices $j$ ($j=f_1NN,\eta NN$) are taken in the
form~(\ref{monFF}).

In the case of $\eta$ exchange we use $g_{\eta NN}=6.1$,
$\Lambda_{\eta NN}$=1.5 GeV from \cite{Holinde} and $g_{\eta\pi
a_0}= 2.2$~GeV (see \cite{Kondrat02}). The contribution of the
$f_1$ exchange is calculated using $g_{f_1 NN}=14.6$,
$\Lambda_{f_1 NN}=2$~GeV from \cite{Kirchbach} and
$g_{f_1a_0\pi}$=2.5. The latter value for $g_{f_1 a_0 \pi}$
corresponds  to $\Gamma(f_1\rightarrow a_0\pi)=24$~MeV and ${\rm
Br}(f_1\rightarrow a_0\pi)=34\%$ (see Ref.~\cite{PDG}).
Eq.(\ref{eq2}) as well as Eq.(\ref{eq3}) can be represented in the
form~(\ref{helamp}) with the invariant amplitudes $A(s,t)$ and
$B(s,t)$ given by
\begin{eqnarray}
&& A^{\{\eta \}}(s,t)= - g_{\eta\pi a_0} g_{\eta NN}\
\frac{F_{\eta\pi a_0}(t) F_{\eta NN}(t)}{t-m_\eta ^2} \ ,\nonumber
\\
&& B^{\{\eta \}}(s,t)= 0
 \end{eqnarray}
for the $\eta$-exchange contribution and
\begin{eqnarray}
 && A^{\{f_1 \}}(s,t,u)= 2 m_N \ \frac{s+t+u-2(m_{a_0}^2+m_N^2)}{m_{f_1}^2}
\nonumber \\ &&
\times g_{f_1\pi a_0} g_{f_1NN}\
 \frac{F_{f_1\pi a_0}(t) F_{f_1NN}(t)}{t-m_{f_1}^2}
 \ , \nonumber
\\
 && B^{\{f_1 \}}(s,t)=2 \ g_{f_1\pi a_0} g_{f_1NN}\
 \frac{F_{f_1\pi a_0}(t) F_{f_1NN}(t)}{t-m_{f_1}^2}
\end{eqnarray}
for the $f_1$-exchange.

The amplitudes of the $s$- and $u$-channel contributions are
defined by the standard expressions:
\begin{eqnarray}
&&\hspace*{-4mm}M_N^s(\pi^-p\rightarrow a_0^0n) = - \sqrt{2}\
g_{a_0NN} {f_{\pi NN}\over m_\pi} \ {1\over s-m_N^2} F_N(s)
\nonumber
\\ &\times&p_{1\mu}\ \bar u(p_2^\prime)  \left[(p_1+p_2)_\alpha
\gamma_\alpha +m_N\right] \gamma_\mu \ \gamma_5 u(p_2);
\label{eqpip1}
\end{eqnarray}
\begin{eqnarray}
&&\hspace*{-4mm}M_N^u(\pi^-p\rightarrow a_0^0n) = \sqrt{2}\
g_{a_0NN} {f_{\pi NN}\over m_\pi} \ {1\over u-m_N^2} F_N(u)
 \nonumber\\ &\times&p_{1\mu} \ \bar u(p_2^\prime)
\gamma_\mu \gamma_5 \left[(p_2-p_1^\prime)_\alpha \gamma_\alpha +
m_N\right] u(p_2), \label{eqpip2}
\end{eqnarray}
where $s=(p_1+p_2)^2, \ u=(p_2-p_1^\prime)^2$, $m_N$ is the
nucleon mass, $f_{\pi NN}^2/4\pi =0.08$~\cite{Holinde}. The form
factor for a virtual nucleon is taken as
\begin{eqnarray}
F_N(u) =
\left(\frac{\Lambda_N^4}{\Lambda_N^4+(u-m_N^2)^2}\right)^j \ ,
\label{FN}
\end{eqnarray}
where~$j=2$, $\Lambda_N $ is the cut-off parameter chosen as
$\Lambda_N = 1.3$~GeV. In Ref.~\cite{Kondrat02} it was found that
the $u$-channel $a_0$ resonance contribution to the $\pi^+
p\rightarrow p K^+ \bar{K}^0$ reaction calculated with the nucleon
form factor $F_N(u)$~(\ref{FN}) of dipole type (j=2) with
$\Lambda_{N}\leq 1.35$~GeV is in a reasonable agreement with
existing experimental data.

Coming back to the amplitudes $A(s,t)$
and $B(s,t)$ defined by Eq.~(\ref{helamp}) we find
\begin{eqnarray}
&& A^{\{s \}}(s,t)= \sqrt{2}\ (s+m_{N}^2) \ g_{a_0 NN}
\frac{f_{\pi NN}}{m_{\pi}} \ \frac{F_N(s)}{s-m_N^2} \ ,\nonumber
\\
 && B^{\{s \}}(s,t)= -\sqrt{2}\ 2m_{N} \
 g_{a_0 NN} \frac{f_{\pi NN}}{m_{\pi}} \  \frac{F_N(s)}{s-m_N^2}
 \end{eqnarray}
for the $s$-channel contribution and
\begin{eqnarray}
 && A^{\{u \}}(s,u)= -\sqrt{2}\ (u+m_{N}^2) \ g_{a_0 NN}
\frac{f_{\pi NN}}{m_{\pi}} \ \frac{F_N(u)}{u-m_N^2} \ , \nonumber
\\
 && B^{\{u \}}(s,u)= \sqrt{2}\ 2m_{N} \
 g_{a_0 NN} \frac{f_{\pi NN}}{m_{\pi}} \  \frac{F_N(u)}{u-m_N^2}
\end{eqnarray}
in the case of the $u$-channel mechanism.

 In the case of the Regge-pole model with the $\rho_2$ and $b_1$ exchanges
we have used the parametrization for $A(s,t)$ and $B(s,t)$ as suggested by
Achasov and Shestakov~\cite{Achasov2}
\begin{eqnarray}
 A^{\{\mbox{Regge}\}}(s,t) \approx
 \frac{\gamma_{b_1}(t)}{\sqrt{s_0}} i\exp \left[-i
{\pi\over 2} \alpha_{b_1}(t)\right] \left(\frac
s{s_0}\right)^{\alpha_{b_1}(t)} \hspace*{-6mm}  ,
\hspace*{6mm}\label{Reg4}
\end{eqnarray}
\begin{eqnarray}
B^{\{\mbox{Regge}\}}(s,t)  &\approx& -\frac{\gamma_{\rho_2}(t)}{s} \exp \
\left[ {-i} \frac{\pi}{2} \alpha_{\rho _2}(t)\right] \left( \frac
s{s_0}\right)^{\alpha_{\rho_2}(t)}\hspace*{-6mm}  , \hspace*{6mm}
\label{Reg3}
\end{eqnarray}
where $$\gamma _{\rho_2}(t) = \gamma_{\rho _2}(0)\ \exp
(b_{\rho_2}t),$$ $$\gamma_{b_1}(t)\ = \gamma_{b_1}(0)\ \exp
(b_{b_1}t),$$ and $s_0\approx 1$ GeV$^2$. The meson Regge trajectories were taken in
the linear form $\alpha_j(t) = \alpha_j(0)+\alpha_j^\prime(0)t$.
The parameters of the residues $\gamma_{\rho_2}(0)$, $b_{\rho_2}$
and $\gamma_{b_1}(0)$, $b_{b_1}$ were fixed in Ref.~\cite{Kondrat02}
using the Achasov and Shestakov fit of the Brookhaven data
on the $\pi^{-} p \rightarrow a_0^{0} n$ reaction at
18~GeV/c~\cite{Dzierba}. They found two solutions with the
relative $b_1$ contribution equal to 0 (fit 1) and 30\% (fit 2).
We use these two different choices of the Regge model for the
analysis of the $\pi N \rightarrow a_0 N$ reaction.

\subsection*{Acknowledgements}
We are very grateful to C. Hanhart for many useful discussions
and clarifying remarks.
This work was supported by DFG (grant 436 RUS 113/630) and
by Russian Fund for Basic Research (grants 02-02-16783 and
03-02-04025). 


\end{document}